%% file: main.tex
\documentclass[fleqn,11pt]{wlscirep}
\usepackage[utf8]{inputenc}
\usepackage[T1]{fontenc}
\usepackage{lineno}
\usepackage[version=4]{mhchem}
\usepackage{textcomp, gensymb}
\usepackage{xr}
\usepackage{tabularx}   
\usepackage{booktabs}   
\usepackage{ragged2e}   
\usepackage{enumitem}   
\title{Autonomous battery research: Principles of heuristic \textit{operando} experimentation}

\author[1,2,3,$\dag$]{Emily Lu}
\author[1,3]{Gabriel Perez}
\author[1,3]{Peter Baker}
\author[4]{Daniel Irving}
\author[4]{Santosh Kumar}
\author[4]{Veronica Celorrio}
\author[1]{Sylvia Britto}
\author[1]{Thomas F. Headen}
\author[4]{Miguel Gomez-Gonzalez}
\author[3,5]{Connor Wright}
\author[4,5]{Calum Green}
\author[1]{Robert Scott Young}
\author[1]{Oleg Kirichek}
\author[1]{Ali Mortazavi}
\author[4]{Sarah Day}
\author[1,3,4,6,7]{Isabel Antony}
\author[1]{Zoe Wright}
\author[1]{Thomas Wood}
\author[4]{Tim Snow}
\author[8]{Jeyan Thiyagalingam}
\author[9]{Paul Quinn}
\author[1]{Martin Owen Jones}
\author[1]{William David}
\author[1,3,10,*,$\dag$]{James Le Houx}

\affil[1]{ISIS Neutron \& Muon Source, Rutherford Appleton Laboratory, Didcot, OX11 0QX, United Kingdom}
\affil[2]{University of Cambridge, The Old Schools, Trinity Ln, Cambridge CB2 1TN, United Kingdom}
\affil[3]{The Faraday Institution, Harwell Science and Innovation Campus, Didcot, OX11 0RA, United Kingdom}
\affil[4]{Diamond Light Source, Rutherford Appleton Laboratory, Didcot, OX11 0QX, United Kingdom}
\affil[5]{Imperial College London, Department of Mechanical Engineering, London, SW7 2AZ, United Kingdom}
\affil[6]{University College London, 117 Gower Street, London WC1E 6AP}
\affil[7]{University of Oxford, Wellington Square, Oxford OX1 2JD}
\affil[8]{Scientific Computing Department, Rutherford Appleton Laboratory, Didcot, OX11 0QX, United Kingdom}
\affil[9]{Ada Lovelace Centre, Rutherford Appleton Laboratory, Didcot, OX11 0QX, United Kingdom}
\affil[10]{University of Greenwich, Old Royal Naval College, Park Row, London, SE10 9LS, United Kingdom}

\affil[$\dag$]{these authors contributed equally to this work}
\affil[*]{Corresponding author: James Le Houx (james.le-houx@stfc.ac.uk)}

\input{Text/01_Abstract}

\begin{document}
\flushbottom
\maketitle

\thispagestyle{empty}
\input{Text/02_Introduction}
\input{Text/03_Results}

\input{Text/Review}
\input{Text/04_Discussion}

\input{Text/05_Conclusion}
\input{Text/06_Methods}
\section*{Acknowledgements}
J.L.H. was partially funded by the Rutherford Appleton Laboratory and the Faraday Institution through his Emerging Leader Fellowship (FIELF001), and by Research England's 'Expanding Excellence in England' grant through the “Multi-scale Multi-disciplinary Modelling for Impact” program ($M^3$4Impact). E.L. was supported by a Faraday Undergraduate Summer Experience (FUSE) studentship (FITG-FUSE228). The authors thank the staff at the ISIS Neutron and Muon Source and Diamond Light Source for their technical support and access to facilities.

\section*{Author contributions statement}
J.L.H. led the Conceptualisation, developed the Heuristic \textit{Operando} Methodology, and was responsible for Funding acquisition and Supervision. E.L. performed the Formal analysis through the hardware review and 3R benchmarking analysis. G.P. provided Supervision and Investigation as the secondary project supervisor. P.B., V.C., D.I., S.K., S.B., T.F.H., M.G.G., C.W., O.K., A.M., I.A., Z.W., R.S.Y., and S.D. contributed to Resources and Investigation by designing and developing the multimodal electrochemical cells evaluated in this study. C.G., T.S., P.Q., J.T., M.O.J., B.D., and T.W. provided Validation and Writing review and editing, specifically regarding the AI framework, real-time data pipelines, and strategic framing of the scientific metrics. J.L.H. and E.L. wrote the original draft. All authors contributed to the technical validation of instrument capabilities and approved the final version.

\section*{Competing interests}
The authors declare no competing interests.

\bibliography{sample}

\end{document}

%% file: Text/01_Abstract.tex
\begin{abstract}
Unravelling the complex processes governing battery degradation is critical to the energy transition, yet the efficacy of \textit{operando} characterisation is severely constrained by a lack of Reliability, Representativeness, and Reproducibility (the 3Rs). Current methods rely on bespoke hardware and passive, pre-programmed methodologies that are ill-equipped to capture stochastic failure events. Here, using the Rutherford Appleton Laboratory’s multi-modal toolkit as a case study, we expose the systemic inability of conventional experiments to capture transient phenomena like dendrite initiation. To address this, we propose Heuristic \textit{Operando} experiments: a framework where an AI pilot leverages physics-based digital twins to actively steer the beamline to predict and deterministically capture these rare events. Distinct from uncertainty-driven active learning, this proactive search anticipates failure precursors, redefining experimental efficiency via an entropy-based metric that prioritises scientific insight per photon, neutron, or muon. By focusing measurements only on mechanistically decisive moments, this framework simultaneously mitigates beam damage and drastically reduces data redundancy. When integrated with FAIR data principles, this approach serves as a blueprint for the trusted autonomous battery laboratories of the future.\\

\textbf{Keywords:} Autonomous Experimentation, Operando Characterisation, Battery degradation, Scientific AI, Physics-Informed Machine Learning

\end{abstract}

%% file: Text/02_Introduction.tex
\section*{Introduction}

Next-generation energy storage technologies, particularly batteries, are indispensable for the global transition to clean energy \cite{grey2017sustainability}. Realising their full potential, however, is hindered by a fundamental bottleneck: the need to understand and mitigate the complex physical and chemical processes, many of them transient and stochastic, that govern performance, safety, and lifecycle. \textit{Operando} characterisation, which probes battery materials during live electrochemical cycling, has become the primary tool for unravelling these dynamics \cite{ziesche2023multi}. Nevertheless, this vital capability is severely constrained. A recent comment by Drnec and Lyonnard \cite{drnec} formalised widespread community concerns, identifying systemic failures in Reliability, Representativeness, and Reproducibility (the 3Rs) as a grand challenge for the entire battery characterisation field. Resolving the 3R crisis extends far beyond the laboratory; it is the rate-limiting step in deploying the safe, next-generation storage technologies essential for a sustainable future.

The urgency of the 3R challenge is compounded by concurrent shifts in experimental capabilities. Next-generation synchrotron and neutron sources, such as the Diamond-II upgrade \cite{chapon2019diamond}, are designed for significantly higher data acquisition rates. Without corresponding methodological advancements, this increased data throughput will inevitably exacerbate the 'data deluge', overwhelming researchers with vast quantities of data that are difficult to store and analyse, while compounding existing difficulties in obtaining reproducible results \cite{bicer2017real}. Simultaneously, recent advances in predictive AI algorithms, the advent of exascale computing \cite{messina2017exascale} and the increased accessibility of high-performance computing (HPC) provide the necessary tools for real-time processing of complex data streams. The conventional, brute-force approach of passive data collection is not only insensitive to transient events but often logistically impossible. As recent work by Corrao et al. \cite{corrao2025modular} highlights, a full, multi-modal map of a single sample wafer could take 2 days by XRD but a staggering 4 months by XAFS. This unsustainable reality demands a methodological shift from passive data collection to an active, intelligent search. Together, these developments provide the technical motivation for fundamentally redesigning \textit{operando} battery characterisation.

In response to the 3R challenge, this paper first provides a critical review of the state-of-the-art, multi-modal operando electrochemical toolkit at the Rutherford Appleton Laboratory (RAL). This analysis, detailed in the Supplementary Information, confirms that the difficulties in achieving the 3Rs are often systemic and linked to the inherent limitations of current, highly specialised hardware. Notably, while specialised cells are highly representative of fundamental material electrochemistry (low Technology Readiness Level (TRL)), they often fail to represent the coupled thermal and mechanical constraints of commercial devices (high TRL). However, our analysis also reveals a deeper methodological problem extending beyond hardware standardisation. We argue that standardisation, while necessary, is an insufficient solution because the fundamental experimental approach itself is flawed.

Conventional \textit{operando} experiments typically rely on pre-programmed, fixed-cadence measurements, rendering them inherently passive. As studies on dynamic processes have shown, this creates a fundamental trade-off: fast scans suffer from poor signal-to-noise, while slow scans lack the temporal resolution to capture short-lived intermediates \cite{szymanski2023adaptively}. Consequently, this approach is blind to unpredictable, stochastic events—such as the nucleation of new phases, the onset of critical interfacial changes (e.g., SEI breakdown), or transient shifts in local ion dynamics—that frequently govern irreversible performance loss and degradation \cite{edge2021lithium}. By definition, a pre-programmed scan is not synchronised with the unpredictable event; it is statistically likely to miss the critical onset of failure. While the burgeoning field of autonomous experimentation (AE) offers potential solutions \cite{stach2021autonomous}, current approaches predominantly focus on active learning strategies, often using Bayesian optimisation or Gaussian processes to efficiently map static parameter spaces \cite{mcdannald2022fly} or optimise material properties \cite{burger2020mobile}. These methods excel at \textit{reactive} optimisation—learning from what has already happened. However, they are fundamentally ill-suited to the challenge addressed here: proactively predicting and intercepting rare, non-stationary, stochastic events governed by hidden chemo-mechanical precursors.

%% file: Text/03_Results.tex
\section*{A Multi-Modal, Multi-Scale Toolkit}

\subsection*{Complementary Probes for a Complex System}

Understanding battery performance and degradation requires probing complex, coupled processes that span multiple length and time scales, from atomic arrangements to electrode-level heterogeneities. Consequently, no single experimental probe can comprehensively capture this intricate behaviour. A multi-modal approach is therefore essential, leveraging the complementary strengths of techniques available at large-scale facilities \cite{ziesche2023multi}. X-ray methods provide high-resolution structural information—ranging from diffraction (XRD) for long-range crystal order to pair distribution function analysis (XPDF) for local, disordered structures. Spectroscopy techniques provide chemical information, such as average bulk oxidation states from X-ray Absorption Spectroscopy (XAS) and surface composition from X-ray Photoelectron Spectroscopy (XPS) \cite{lin}. Neutron techniques offer complementary contrast because, unlike X-rays, neutrons interact with the atomic nucleus \cite{headen}. This fundamental difference provides unique capabilities, including sensitivity to light elements (like Li and H), the ability to distinguish between isotopes (e.g., ${}^{6}Li/^{7}Li$), and the ability to differentiate elements with similar electron configurations, such as neighboring transition metals \cite{perez}. Muon spectroscopy ($\mu$SR) is used as a sensitive local probe capable of quantifying ion diffusion dynamics, offering atomic-level insights complementary to neutron and X-ray methods \cite{blundell}. Electron microscopy (SEM/TEM) provides unparalleled spatial resolution for micro- and nanostructure, while laser-based vibrational spectroscopy (e.g., Raman, IR) and neutron molecular spectroscopy adds chemical sensitivity, probing molecular bonds and interfacial species.
Integrating these diverse techniques is vital for a holistic understanding. However, this necessity inherently drives the requirement for a varied and complex suite of specialised \textit{operando} electrochemical cells, presenting significant practical challenges.

\subsubsection*{X-ray techniques}
Synchrotron facilities provide high-brilliance, tunable-energy X-ray beams essential for probing battery materials. The wide energy range, from soft X-rays (typically 0.1–2 keV) to high-energy hard X-rays (exceeding 100 keV), enables access to complementary structural and chemical information across different probing depths and buried interfaces. While higher energies offer greater penetration due to minimal interaction with matter, the intense photon flux required for time-resolved experiments significantly increases the risk of beam-induced degradation (e.g., electrolyte radiolysis). This trade-off necessitates careful beam modulation to ensure data \textit{Reliability}. Nonetheless, X-rays remain indispensable.

Key X-ray methods applied to batteries include scattering, spectroscopy, and imaging \cite{lin}. X-ray diffraction (XRD) is widely used to determine crystal structure, lattice parameters, strain, and phase evolution during electrochemical cycling. X-ray Pair Distribution Function (XPDF) analysis extends structural insights to materials lacking long-range order, making it well-suited for studying nanocrystalline, amorphous, or liquid components. Spectroscopic techniques like X-ray Absorption Spectroscopy (XAS) provide information on oxidation states, bond lengths, and local coordination environments via XANES and EXAFS analysis \cite{lin}. Resonant Inelastic X-ray Scattering (RIXS) offers complementary electronic structure information, probing orbital states and charge transfer dynamics. X-ray Photoelectron Spectroscopy (XPS) offers surface sensitivity, yielding elemental and chemical information important for characterising interfaces like the solid-electrolyte interphase (SEI). Finally, X-ray Computed Tomography (XCT) provides 3D morphological information, enabling visualisation of electrode microstructures, particle cracking, gas evolution, and dendrite formation \cite{le2021x}. More advanced methods are also emerging: ptychography enables high-resolution phase-contrast imaging for nanoscale morphology; XRD-CT combines diffraction contrast with tomography to map phase distributions and strain fields in 3D; and Dark Field X-ray Microscopy (DFXM) allows high-resolution mapping of crystal orientation and strain within individual grains \cite{le2025nanoscale}. Each technique often requires specific operando cell designs tailored to its geometry and experimental requirements.

\subsubsection*{Neutron techniques}
Neutron analysis techniques provide complementary information to X-rays \cite{headen}. Unlike X-rays, which interact primarily with the electron cloud, neutrons interact with the atomic nucleus. This fundamental difference grants them unique sensitivity to light elements like lithium and hydrogen, the ability to distinguish isotopes (e.g., $^6$Li/$^7$Li), and the ability to differentiate between elements with similar electron configurations, such as neighboring transition metals. These properties make neutrons particularly well-suited for probing lithium distribution and transport within battery materials \textit{operando} \cite{perez}. Isotopic substitution further enhances contrast; common examples include using $^{6}$Li/$^{7}$Li labelling or deuteration (replacing H with D). While often used to minimise the incoherent scattering background from hydrogen, this technique is also critical in neutron total scattering to elucidate the atomic-scale structure of hydrogen-containing liquid electrolytes. For example, it has recently been applied to resolve lithium solvation shells and anion-dominated domains in water-in-salt (WIS) systems \cite{groves2025lithium}. Due to their weak interaction with matter, neutrons are highly penetrating and cause negligible radiation damage, enabling bulk studies. Key neutron methods probe a wide range of properties. Structural techniques include neutron diffraction (ND) for determining long-range crystal structures; Neutron Total Scattering (NTS) for characterising non-crystalline and disordered materials, ranging from solids (e.g., hard carbon) to liquids (e.g., electrolytes); Small Angle Neutron Scattering (SANS) for probing nanoscale structures like porosity; and neutron reflectometry (NR) for measuring thin films and buried interfaces. Neutron spectroscopy is used to investigate material dynamics, such as Inelastic Neutron Scattering (INS) for vibrational modes and Quasi-Elastic Neutron Scattering (QENS) for slower diffusional processes like Li-ion hopping. Finally, imaging techniques, including radiography, tomography, and Bragg edge imaging, are used to visualise features such as gas evolution, Li distribution, and strain fields. Collectively, these neutron techniques provide fundamental insights into the structural, morphological, and dynamic behaviour of battery components, often inaccessible by other means.
\subsubsection*{Muon techniques}
Muon Spectroscopy ($\mu$SR) offers a sensitive bulk probe of the local atomic environment within materials \cite{blundell}. By implanting spin-polarized muons (either positive, $\mu^+$, or negative, $\mu^-$) into a sample and detecting their decay products (positrons or electrons), $\mu$SR measures the local magnetic fields experienced by the muon \cite{mcclelland}. This makes it a powerful tool for studying magnetism, superconductivity, and charge transport. In battery research, $\mu^+$SR is particularly valuable for quantifying ion diffusion rates and pathways, especially for species like Li$^+$ and Na$^+$ which possess suitable nuclear magnetic moments. Such experiments can also be performed with negative muons ($\mu^-$SR) to confirm that the muon remains static. 
Complementarily, negative muons can also be used for elemental analysis in a technique called $\mu$XES (muon X-ray Emission Spectroscopy), where an implanted negative muon ($\mu^-$) is captured by an atom, forming an unstable muonic atom. The muon then cascades down through its orbital energy levels, emitting characteristic muonic X-rays as it transitions. Because the energy of these X-rays is specific to the nuclear charge (Z) of the capturing element, this method provides a sensitive, non-destructive probe for elemental composition that is sensitive far below the surface of the sample. The technique provides atomic-level insights complementary to neutron and X-ray methods.
\subsubsection*{Electron techniques}
Electron Microscopy (EM) techniques, such as Scanning Electron Microscopy (SEM) and Transmission Electron Microscopy (TEM), provide unparalleled spatial resolution for visualising battery micro- and nanostructures. SEM is widely used for imaging electrode surfaces and morphology, while TEM can resolve atomic structures within materials. Focused Ion Beam SEM (FIB-SEM) enables 3D reconstruction of electrode volumes. Coupled with Energy-Dispersive X-ray Spectroscopy (EDS) or Electron Energy Loss Spectroscopy (EELS), EM also provides localised elemental and chemical information. However, EM typically requires vacuum conditions and specific sample preparation (e.g., thin sectioning for TEM), often limiting \textit{operando} studies to specialised setups or ex-situ analysis.
\subsubsection*{Laser-based techniques}
Laser-based methods provide valuable complementary chemical and structural information, often suitable for \textit{in-situ} or \textit{operando} measurements. Vibrational spectroscopy, notably Raman and infrared (IR) spectroscopy, probes molecular bonds and lattice vibrations, offering sensitivity to crystal phases, electrolyte composition, and the formation of interfacial species like the solid-electrolyte interphase (SEI). Ultrafast laser techniques can investigate charge carrier dynamics and energy transfer processes on femtosecond to picosecond timescales. Furthermore, advanced laser-based microscopy approaches enable high-resolution chemical imaging. While powerful, challenges can include sample fluorescence, potential laser-induced heating, and penetration depth limitations compared to X-ray or neutron probes.

\subsection*{Reviewing \textit{Operando} Hardware Trade-offs}

Each of the complementary techniques described above typically requires a unique \textit{operando} electrochemical cell, designed with specific geometries, window materials, and operational constraints tailored to the experimental method. A detailed technical review of each major cell type employed within the RAL multi-modal toolkit, including the POLARIS cell for neutron diffraction, the BAM cell for muon spectroscopy, the M-series cells for nano-focus X-ray spectroscopy, the DRIX cell for X-ray PDF, and various flow and static cells for X-ray spectroscopy, is provided in the Supplementary Information. To synthesise the key findings relevant to the challenges of reproducibility, representativeness, and reliability, Table \ref{tab:synthesis} summarises the primary limitations identified for each cell design based on this review.

\begin{table}[h!]
  \centering
  \caption{Analysis of inherent trade-offs in specialised \textit{operando} electrochemical cells at RAL, highlighting factors relevant to achieving Reliability, Representativeness, and Reproducibility (3R). See Supplementary Information for detailed cell descriptions.}
  \label{tab:synthesis}
  \small
  \begin{tabularx}{\textwidth}{@{} 
      >{\raggedright\arraybackslash\hsize=0.5\hsize}X 
      >{\raggedright\arraybackslash\hsize=0.8\hsize}X 
      >{\raggedright\arraybackslash\hsize=1.2\hsize}X 
      >{\raggedright\arraybackslash\hsize=1.3\hsize}X 
      >{\raggedright\arraybackslash\hsize=1.2\hsize}X 
  @{}}
    \toprule
    \textbf{Cell Name} & \textbf{Primary Technique(s)} & \textbf{Reliability Issues} & \textbf{Representativeness Issues} & \textbf{Reproducibility Issues} \\
    \midrule

    \textbf{POLARIS} & Neutron Diffraction &
      \begin{itemize}[nosep, leftmargin=*]
        \item Unquantified/ inconsistent stack pressure
        \item Electrolyte settling/ heterogeneity 
      \end{itemize} &
      \begin{itemize}[nosep, leftmargin=*]
        \item Low TRL: Representative of materials, not commercial devices
        \item High mass loading affects ion diffusion
      \end{itemize} &
      \begin{itemize}[nosep, leftmargin=*]
        \item Non-standard, in-house design
        \item Pressure inconsistency from PEEK screws
      \end{itemize} \\
    \addlinespace

    \textbf{BAM} & Muon Spectroscopy ($\mu$SR), SANS, Neutron Diffraction &
      \begin{itemize}[nosep, leftmargin=*]
        \item Unquantified/ inconsistent stack pressure
        \item Electrolyte settling/ heterogeneity
      \end{itemize} &
      \begin{itemize}[nosep, leftmargin=*]
        \item Low TRL: Representative of materials, not commercial devices
        \item High mass loading limits C-rate ($C/5$)
      \end{itemize} &
      \begin{itemize}[nosep, leftmargin=*]
        \item Non-standard, in-house design
        \item Requires user balance: electrochem vs. data
      \end{itemize} \\
    \addlinespace

    \textbf{DRIX} & X-ray PDF &
      \begin{itemize}[nosep, leftmargin=*]
        \item Background variation if beam misaligned or optics not optimised
        \item Fragile capillaries
      \end{itemize} &
       \begin{itemize}[nosep, leftmargin=*]
        \item Low TRL: Representative of materials, not commercial devices
        \item Small sample size
      \end{itemize} &
      \begin{itemize}[nosep, leftmargin=*]
        \item In-house assembly of commercial off-the-shelf (COTS) components
        \item Difficult assembly
        \item Unsuited to pressure
       \end{itemize} \\
     \addlinespace

     \textbf{B07 Flow Cell} & Soft X-ray (XPS, XAS) &
       \begin{itemize}[nosep, leftmargin=*]
         \item Beam damage artifacts
         \item Flow-induced vibrations
         \item Poor S/N if sample too thick
       \end{itemize} &
       \begin{itemize}[nosep, leftmargin=*]
         \item Low TRL: Representative of materials, not commercial devices
         \item Thin sample requirement
       \end{itemize} &
       \begin{itemize}[nosep, leftmargin=*]
         \item Non-standard, in-house design
         \item Membrane prep difficulty/cost
         \item Requires flow optimisation
       \end{itemize} \\
     \addlinespace

     \textbf{B18 Cells} & Hard X-ray (XAS) &
       \begin{itemize}[nosep, leftmargin=*]
         \item Beam damage/ radiolysis risk
         \item Window damage risk
       \end{itemize} &
       \begin{itemize}[nosep, leftmargin=*]
         \item Low TRL: Representative of materials, not commercial devices
       \end{itemize} &
       \begin{itemize}[nosep, leftmargin=*]
         \item Non-standard, in-house designs
       \end{itemize} \\

    \addlinespace
    \textbf{M3.0 / M4.0} & Nano-focus X-ray (XRF, XANES, XRD) &
      \begin{itemize}[nosep, leftmargin=*]
        \item M3.0: Window bulging affects pressure uniformity
        \item M4.0: Risk of electrical shorting from Al plates
      \end{itemize} &
      \begin{itemize}[nosep, leftmargin=*]
        \item (M4.0): Designed for reproducible stack pressure
        \item Limited by small window size
      \end{itemize} &
      \begin{itemize}[nosep, leftmargin=*]
        \item M3.0: Trade-off between stack uniformity and pressure
        \item M4.0: Limited availability of FFKM gaskets
      \end{itemize} \\

    \bottomrule
  \end{tabularx}
\end{table}

%% file: Text/Review.tex
\section*{The 3Rs and the Limits of Conventional Operando Methodology}

The synthesis presented in Table \ref{tab:synthesis} links the design constraints of current \textit{operando} electrochemical cells directly to the 3R challenge identified by Drnec and Lyonnard \cite{drnec}. The difficulties encountered are systemic and hardware-based. For instance, the widespread use of unquantified or inconsistent stack pressures in cells like POLARIS and BAM, or the potential for beam-induced artifacts noted particularly for the B18 and B07 Flow Cells, directly undermines experimental \textit{Reliability}. Similarly, the prevalence of low TRL designs (e.g., DRIX) featuring non-standard geometries reframes the challenge of \textit{Representativeness}: while effective for answering fundamental questions about intrinsic material properties, their findings cannot necessarily be extrapolated to commercial devices.

This systemic lack of representativeness has led to a corresponding push to use commercial cell formats (e.g., coin or pouch cells). However, this approach presents its own critical trade-offs. Many commercial cells are fundamentally incompatible with most probes due to their metallic casings, forcing researchers into a difficult compromise (\textbf{Figure \ref{fig:radar}a}). One path is to modify the commercial cells with windows, which in turn compromises the standardisation and representativeness they were chosen for. The other path is to use unmodified commercial cells directly; while perfectly representative, these introduce severe data quality challenges, such as high signal noise in neutron beamlines or requiring highly specialised techniques like X-ray laminography.

This central dilemma, where specialised cells have 3R flaws, modified commercial cells re-introduce 3R flaws, and unmodified commercial cells create major data quality issues, confirms that no current hardware provides a perfect solution. It underscores that while hardware standardisation is necessary, it is an insufficient solution. Resolving this impasse requires a fundamental methodological shift. We propose the Heuristic \textit{Operando} Framework to directly mitigate the 3R limitations inherent in the specialised hardware used to acquire high-quality data.

Current experimental strategies fail to address this gap because they typically rely on passive, pre-programmed, fixed-cadence data acquisition. This approach is not only inefficient, generating vast quantities of scientifically mundane data in a brute-force attempt to capture rare events and thus contributing significantly to the data deluge faced by large-scale facilities, but it is also inherently blind to the unpredictable, stochastic events that often initiate irreversible performance degradation or failure. Consequently, the true bottleneck lies not only in the quality of data collected (the 3Rs) but in the fundamental inability of predefined protocols to capture crucial, fleeting phenomena, such as the nucleation of detrimental phases, the initiation of dendrite growth, or the transient chemo-mechanical stresses preceding particle fracture, precisely when and where they occur.

%% file: Text/04_Discussion.tex
\section*{Heuristic \textit{Operando} Experimentation}

\subsection*{From Passive Data Collection to Active Scientific Search}
\begin{figure}[h!]
    \centering
    \includegraphics[scale = 1.2]{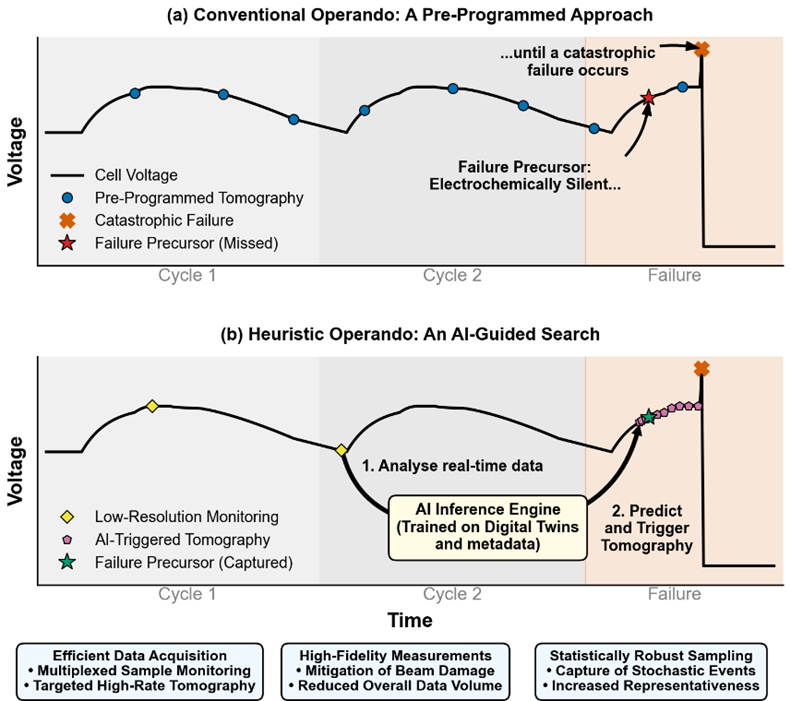}
    \caption{\textbf{Capturing stochastic events: Conventional vs. Heuristic \textit{Operando.}} (a) Conventional \textit{operando} often misses stochastic failure precursors (red star) that occur between pre-programmed, fixed-cadence scans (blue circles). (b) Heuristic \textit{Operando} uses an AI (trained on physics-based digital twins) to identify precursors in continuous low-resolution monitoring data (yellow diamonds). This detection triggers a targeted, high-cadence scan (pink pentagons), deterministically capturing the precursor event (green star).}
    \label{fig:heuristic}
\end{figure}

To address the limitations of conventional methodologies outlined above, particularly the inability to capture stochastic phenomena, a fundamental shift in experimental strategy is required. We propose Heuristic \textit{Operando} experiments, designed to transform passive, brute-force data collection into an active, intelligent search for scientifically critical events. The core concept is illustrated in \textbf{Figure \ref{fig:heuristic}}. Conventional \textit{operando} experiments, using tomography as an example, (Fig. \ref{fig:heuristic}a) typically employ pre-programmed, fixed-cadence measurements (blue circles). While straightforward to implement, this approach frequently fails to capture transient failure precursors (red star) that occur unpredictably between scheduled scans. In contrast, the Heuristic \textit{Operando} approach (\textbf{Fig. \ref{fig:heuristic}b}) employs an AI Inference Engine (or AI Pilot) to execute an active, intelligent search. To identify the subtle, non-stationary precursor signatures hypothesised to reliably precede critical events, the AI Pilot is trained on libraries of physics-based simulations (digital twins) \cite{le2021openimpala} and prior experimental data. During the experiment, the pilot continuously analyses a stream of monitoring data (yellow diamonds), which may be low-resolution or from a secondary, high-speed probe. Upon detecting a precursor signature, it predicts the spatio-temporal hotspot of the incipient event. This prediction then autonomously triggers a targeted, high-cadence, high-resolution measurement (pink pentagons), focusing the instrument's full capability only on that specific location and moment. This targeted approach enables the deterministic capture of the transient precursor event (green star), a feat that passive, pre-programmed scans would statistically miss.

A primary focus of autonomous experimentation at synchrotrons has been developing active learning frameworks to efficiently map static parameter spaces. Pioneering work at facilities like NSLS-II has demonstrated this capability, using decentralised AI agents to autonomously map combinatorial libraries \cite{maffettone2023self} and sophisticated multi-beamline agents to resolve phase boundaries \cite{corrao2025modular}. These experiments are typically rooted in Gaussian Process modeling, often using Kriging methods to guide measurements towards regions of maximum model uncertainty \cite{noack2019kriging}. While highly effective for optimising the exploration of static parameters, this uncertainty-driven reactive strategy is fundamentally distinct from the challenge addressed here. Heuristic \textit{Operando} experiments are not a mapping tool for static properties, but a proactive hunting tool designed to predict and intercept a future, transient event, like dendrite nucleation, within a single, dynamically evolving system. To formalise this distinction, Table \ref{tab:glossary} contextualises the framework within the broader hierarchy of synchrotron methodologies, highlighting the shift from the reactive optimisation of active learning to the proactive event capture required for failure analysis.

\begin{table*}[h]
\centering
\caption{Hierarchy of Synchrotron Characterisation Strategies: From Ex-situ to Heuristic Operando.}
\label{tab:glossary}
\small
\begin{tabular}{p{0.18\linewidth} p{0.45\linewidth} p{0.30\linewidth}}
\toprule
\textbf{Methodology} & \textbf{Context and Optimisation Objective} & \textbf{Control Strategy} \\
\midrule
\textbf{Ex-situ} & Post-mortem or pristine analysis of materials removed from the electrochemical environment. & \textbf{Static:} Single snapshot of a fixed, equilibrium state. \\
\midrule
\textbf{In-situ} & Analysis within the cell environment, typically under static conditions or during open-circuit relaxation. & \textbf{Quasi-static:} Observation of stability or slow relaxation in environment. \\
\midrule
\textbf{Operando} & Analysis during live electrochemical cycling using pre-determined time intervals. & \textbf{Passive:} Fixed-cadence data collection, statistically likely to miss transient onsets. \\
\midrule
\textbf{Active Learning} & Autonomous loop minimising posterior uncertainty over a static parameter space (e.g., mapping equilibrium phase boundaries). & \textbf{Reactive:} Optimises the exploration of a static parameter space based on prior observations. \\
\midrule
\textbf{Heuristic Operando} & Autonomous loop minimising time-to-detection of non-stationary precursors via physics-informed prediction. & \textbf{Proactive:} Anticipates and intercepts transient, non-stationary events (e.g., dendrite nucleation). \\
\bottomrule
\end{tabular}
\end{table*}

Recent advancements have extended these autonomous capabilities from static mapping to dynamic process characterisation. For instance, convolutional neural networks have been coupled directly to diffractometers to autonomously guide data acquisition \cite{szymanski2023adaptively}. By analysing data in real-time and selectively resampling regions based on model confidence, this adaptive XRD approach has improved the speed and accuracy of phase identification. Notably, this method enabled the successful \textit{in-situ} identification of a short-lived intermediate phase during a solid-state reaction. While this demonstrates the power of integrating ML into the experimental loop, this approach is still fundamentally reactive: it optimises its search based on current model confidence. It is not designed to be proactive; it does not use physics-based precursors to predict the onset of a future, non-stationary event before it begins.

\par

\begin{figure}[h!]
    \centering
    \includegraphics[scale = 0.5]{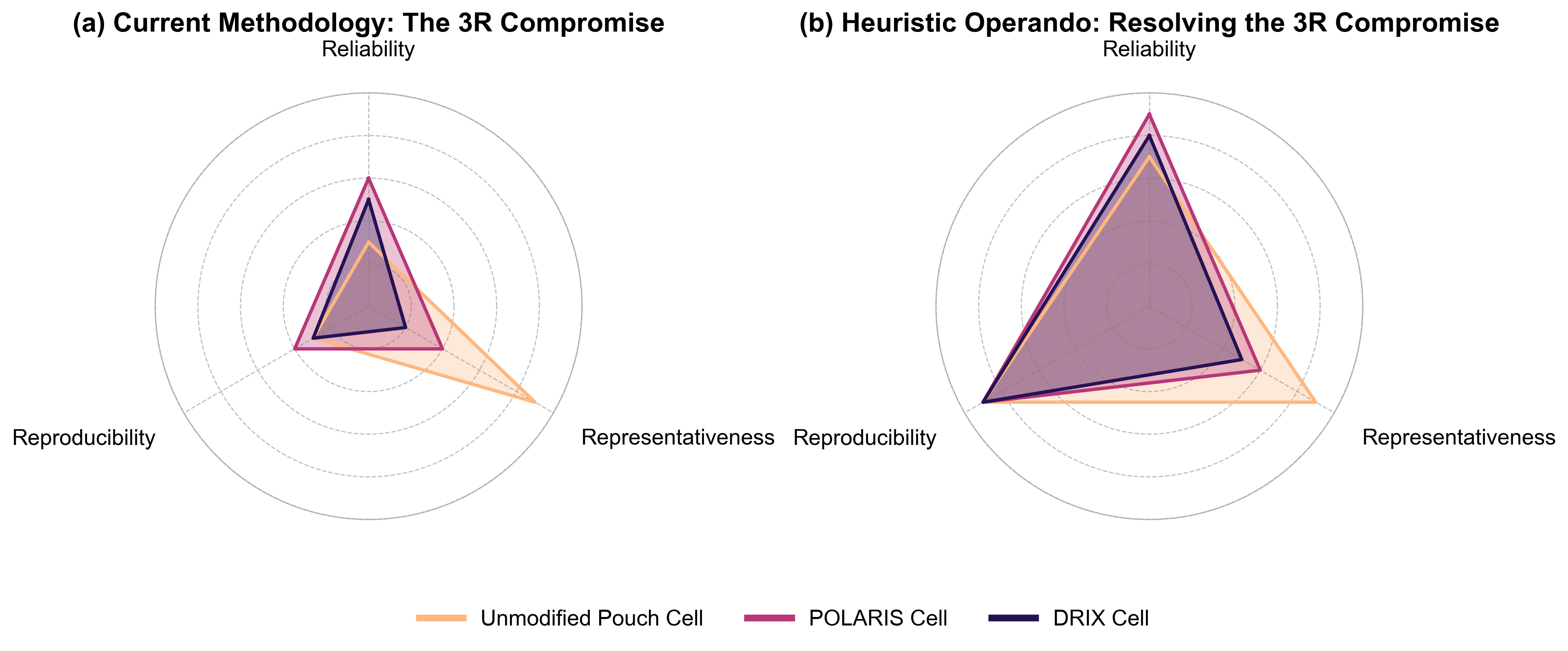}
    \caption{\textbf{Resolving the 3R (Reliability, Representativeness, Reproducibility) Compromise.} (a) \textbf{Current Methodology:} The fundamental hardware trade-off. Researchers are forced to choose between high-Representativeness commercial cells (e.g., Unmodified Pouch), which suffer from poor signal quality (low Reliability), or high-Reliability specialised cells (e.g., POLARIS, DRIX), which lack commercial relevance. (b) \textbf{Heuristic \textit{Operando:}} The proposed framework breaks this compromise. The AI Pilot makes commercial cells viable by extracting faint precursor signals (boosting Reliability) and upgrades specialised cells by enabling efficient multiplexing and targeted scans (boosting Representativeness).}
    \label{fig:radar}
\end{figure}

Compounding this lack of predictive capability is a fundamental trade-off within the experimental hardware itself, as highlighted by the analysis above. As visualised in Figure \ref{fig:radar}a, researchers are forced to choose between highly Representative but low-Reliability commercial cells and high-Reliability but low-Representativeness specialised cells. Heuristic \textit{Operando} experimentation is designed to break this compromise. By using an AI pilot to intelligently guide the experiment, the framework makes commercial cells viable and specialised cells more statistically robust, effectively transforming both into near-ideal tools as shown in Figure \ref{fig:radar}b.

\par

\subsection*{Theoretical Formulation}

To formalise this active search, we propose a shift from volume-based metrics (throughput) to information-based metrics. We define the Entropy-Scaled Measurement Efficiency ($E_{\mathrm{SME}}$), which quantifies the rate of Mutual Information ($I$) gained regarding the target failure mode ($M$) per unit of generalised experimental cost ($C$). Using Shannon's information theory, this is expressed as:

\begin{equation}E_{\mathrm{SME}} = \frac{I(M; D_{t})}{C} = \frac{H(M) - H(M | D_{t})}{C}\label{eq:efficiency}\end{equation}

\noindent where $H(M)$ represents the initial entropy (uncertainty) of the failure mechanism and $H(M | D_{t})$ represents the remaining uncertainty after collecting data $D$ at time $t$. To operationalise this in real-time contexts, the computation of $H$ depends on the available latency. While fully explicit entropy calculations via Bayesian surrogates are feasible for slower processes, high-rate monitoring necessitates operational approximations, such as using the variance of ensemble models or conformal prediction intervals as a proxy for informational uncertainty. The denominator $C$ is a weighted cost function ($\sum w_i c_i$) accounting for beamtime, data storage, and sample damage (dose). 

Importantly, to prevent confirmation bias, where the algorithm ignores unexpected physics, we explicitly define the hypothesis space $M$ as:

\begin{equation}M = \{m_1, m_2, ..., m_k, m_{\emptyset}\}\label{eq:hypothesis}\end{equation}

\noindent where $\{m_1, ..., m_k\}$ represent known failure modes (e.g., dendrites, cracking) and $m_{\emptyset}$ represents a high-entropy anomaly hypothesis. By enforcing a non-vanishing prior on $m_{\emptyset}$ (i.e., Cromwell's rule), the framework prevents the posterior probability of the unknown from asymptotically converging to zero during long periods of nominal operation. This ensures that $E_{\mathrm{SME}}$ remains high when deviations from known models occur, effectively placing a mathematical premium on the discovery of the unknown.

The applicability of this framework is constrained by two fundamental physical assumptions regarding the target failure mode $m_k$. First, the precursor must satisfy the condition of observability, meaning the low-resolution monitoring stream $D_{mon}$ must contain a causal signature distinguishable from noise. Second, the process must exhibit temporal separability, where the time lag $\Delta t$ between the precursor onset and the failure event exceeds the total latency of the experimental control loop. Failure modes that are effectively instantaneous, such as brittle fracture without plastic onset, or those that are silent to the monitoring modality, therefore lie outside the predictive scope of the AI Pilot.

\subsection*{Integrating HPC-Trained Surrogates}

\begin{figure*}[h!]
    \centering
    \includegraphics[width=\textwidth]{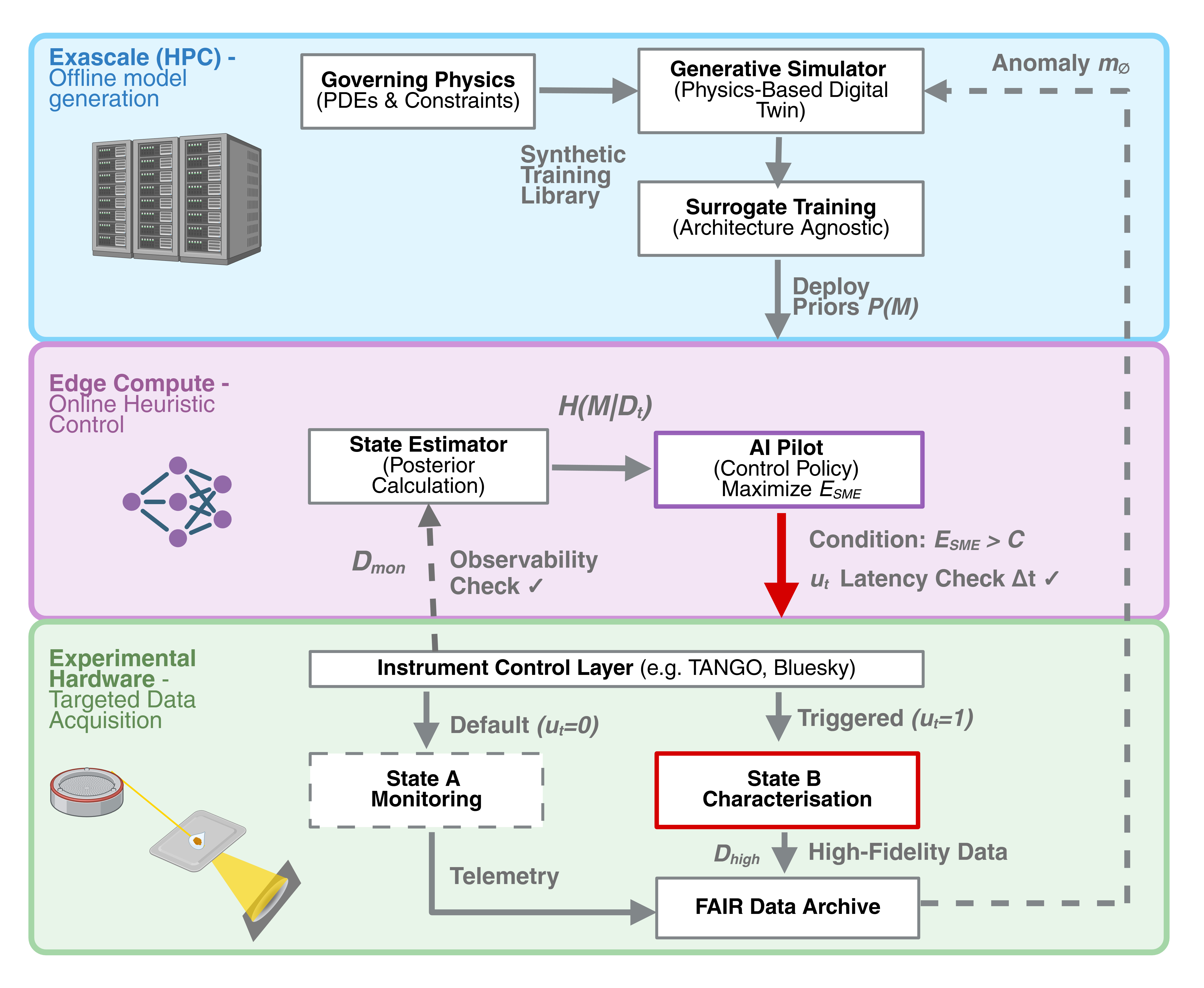}
\caption{\textbf{Closed-loop architecture of the Heuristic Operando Framework.} 
\textbf{Exascale Zone:} Physics-based Digital Twins generate synthetic training libraries for architecture-agnostic surrogates, deploying probabilistic priors ($P(M)$) to the edge.
\textbf{Edge Zone:} The AI Pilot computes the Entropy-Scaled Measurement Efficiency ($E_{\mathrm{SME}}$) from real-time monitoring ($D_{mon}$). A control signal ($u_t$) is triggered only when the information gain exceeds the experimental cost ($C$).
\textbf{Experiment Zone:} The instrument defaults to low-dose monitoring (State A) to mitigate beam damage. Signal $u_t$ activates targeted characterisation (State B), yielding high-fidelity data ($D_{high}$) and identifying anomalies ($m_{\emptyset}$) for model refinement.}
    \label{fig:architecture}
\end{figure*}

To operationalise this framework, we define a strict computational hierarchy that separates physical generation from real-time inference, as illustrated in Figure \ref{fig:architecture}. The architecture is organised into three functional zones. In the Exascale Zone, the Digital Twin serves as the high-fidelity, physics-based generative simulator, utilising solvers like OpenImpala \cite{le2021openimpala} to create an exhaustive offline training library of underlying chemo-mechanical processes. The Surrogate Model acts as an architecture-agnostic, learned, differentiable approximation (e.g., a Neural Operator or Transformer) derived from this library, optimised for millisecond-scale inference and deployed as priors ($P(M)$). In the Edge Zone, the AI Pilot functions as the decision-making agent, querying the Surrogate to forecast system evolution ($m_k$). By training on this physics-based foundation rather than sparse experimental data, the Pilot learns to correlate multi-modal precursors, such as specific electrochemical fingerprints (e.g., soft shorts \cite{menkin2024insights}) or macroscopic strain fields, to anticipate future failure before it becomes observable via conventional thresholds. This predictive logic drives the Experiment Zone, triggering targeted acquisition only when a specific precursor is identified.

To build this predictive capability, the framework adopts an architecture-agnostic approach, where the selection of the specific AI model is governed strictly by the constraints of modality, latency, and governing physics. For instance, while Physics-Informed Neural Operators (PINOs) such as Fourier Neural Operators are well-suited for encoding families of PDEs in diffusion-dominated systems \cite{li2020fourier}, sequence-based architectures like Transformers may be preferable for analysing temporal scattering patterns \cite{yin2024integrated}. Similarly, generative diffusion models offer utility for capturing probabilistic evolution in highly stochastic regimes \cite{du2024conditional, gao2024generative}. Ultimately, the framework requires only that the chosen architecture provides a differentiable, probabilistic forecast of the target state within the latency budget of the control loop. The feasibility of this predictive, HPC-trained approach is already evident in pioneering facility workflows. For example, the edge-to-exascale initiative at the Spallation Neutron Source (SNS) recently employed a Temporal Fusion Transformer, trained on the Frontier supercomputer, to successfully predict the evolution of 3D neutron scattering patterns \cite{yin2024integrated}. Similarly, the Autonomous Neutron Diffraction Explorer (ANDiE) validated the utility of physics-based decision loops by directly encoding Weiss and Ising models to navigate phase transitions \cite{mcdannald2022fly}.

However, reliance on simulation introduces the challenge of the sim-to-real gap. As noted by Maffettone et al., AI agents pre-trained on synthetic data can fail if physical samples are sufficiently out-of-distribution \cite{maffettone2023self}. This raises a fundamental question of circularity: how can we discover new failure mechanisms if the AI is trained only on known physics? The Heuristic approach resolves this through an iterative, spiral discovery process. As formalised in our definition of Entropy-Scaled Measurement Efficiency ($E_{\mathrm{SME}}$), the hypothesis space $M$ is explicitly defined to include an unknown term, $m_{\emptyset}$ (Equation \ref{eq:hypothesis}). This allows the Pilot to function as an anomaly detector: when experimental data significantly diverges from the digital twin's forecast—implying the true state lies within $m_{\emptyset}$—the event is flagged as a potential discovery. This discrepancy data is then captured and fed back to update the digital twin, closing the loop. This establishes a self-reinforcing feedback mechanism where the simulation is progressively refined, enabling the pilot to hunt for increasingly subtle precursors in subsequent experiments.

\subsection*{Hardware and Data Infrastructure}

Implementing the Heuristic \textit{Operando} Framework requires more than just the AI Pilot software; it necessitates a co-designed system integrating specific hardware and data infrastructure components. Key practical requirements include fast detectors and real-time data pipelines capable of acquiring and transferring sufficient data volumes to the AI Pilot with low latency. Robust, modular instrument control software is also essential. This layer accepts and executes high-level commands from the AI Pilot for dynamic experimental steering. Software suites like Bluesky, developed at NSLS-II, are designed precisely for this type of agent-based, event-driven orchestration and provide a clear pathway for implementation. Finally, a Findable, Accessible, Interoperable, and Reusable (FAIR) data platform underpins the entire workflow. This platform serves a dual purpose: providing access to curated historical data for training the AI Pilot, while also archiving both the predicted events and the unanticipated anomalies generated by the heuristic search.

This infrastructure model is already being pioneered at major facilities. The edge-to-exascale workflow at Oak Ridge National Laboratory, for instance, provides a direct blueprint by coupling the Spallation Neutron Source (SNS) with the Frontier exascale supercomputer \cite{yin2024integrated}. This system demonstrates the key hardware components in practice: it employs edge computing (e.g., DGX workstations) at the beamline for rapid data preprocessing and real-time AI inference, while leveraging exascale HPC for large-scale model training. This architecture, part of the Department of Energy's vision for an Integrated Research Infrastructure (IRI), demonstrates the low-latency, HPC-connected system our framework demands. Similarly, work by Pithan et al. at the ESRF explored the practical software integration, connecting ML analysis code via the facility's SCADA system (TANGO controls) or through standardized hardware interfaces like the Dynamo-Triton inference platform \cite{pithan2023closing}. These examples establish a clear pathway for creating the stable, well-isolated software environments needed for both beamline control and user-side AI analysis.

\subsection*{Resolving the 3Rs and Capturing Stochastic Events}

The Heuristic \textit{Operando} Framework is designed to address both the methodological challenge of capturing stochastic events (outlined previously) and the systemic 3R challenges highlighted by our hardware analysis (Table \ref{tab:synthesis}). The framework's primary objective is the characterisation of stochastic phenomena; this is accomplished via the AI Pilot's active, intelligent search, which replaces passive observation with targeted data acquisition. This targeted approach also directly mitigates the impending data deluge by focusing high-resolution measurements only on scientifically critical moments, significantly reducing overall data volume compared to continuous brute-force scanning.

Furthermore, this framework directly enhances data quality in relation to the 3Rs. Reliability is improved by mitigating known hardware limitations. For example, by minimising total exposure time through targeted acquisition, the framework reduces the risk of beam damage artifacts—a noted concern for X-ray techniques (e.g., B07 Flow Cell, B18 Cells). For flux-limited neutron or muon measurements, concentrating measurement time on predicted events enhances statistical accuracy, potentially overcoming limitations associated with low signal-to-noise in passive scans. Reproducibility is fundamentally improved; automating the experimental decision-making process via the AI Pilot creates a precisely defined, software-driven protocol. This removes operator variability in identifying and reacting to transient events, ensuring that the same heuristic strategy can be executed consistently across different experiments, facilities, or user groups.

Finally, the framework fundamentally enhances Representativeness in three key ways, effectively breaking the trade-off visualised in Figure \ref{fig:radar}b. First, it enables statistically robust sampling through multiplexing. By using efficient monitoring and short, targeted measurements, the framework frees up beamtime. This allows multiple, nominally identical cells to be run in parallel, enabling the AI Pilot to hunt for and capture a population of stochastic events. Second, it enables the use of more industrially relevant cells. An AI Pilot, trained to detect faint, time-evolving precursor signatures, could successfully identify an incipient event even within the high-noise data from a fully representative commercial cell—a task impossible for a passive scan. Third, even when using specialised cells, the framework enhances phenomenological representativeness by capturing the true underlying physics of stochastic initiation events, providing data that is more representative of fundamental failure mechanisms.

\subsection*{Foundations of Heuristic Control}

The Heuristic \textit{Operando} Framework, while ambitious, builds upon a growing foundation of AI-driven methodologies. Existing examples serve as important stepping stones. Notably, techniques such as Sparse-XANES for optimising spatial sampling and Super-Resolution GANs for image enhancement are now being executed in real-time alongside data acquisition \cite{de_castro_vargas_fernandes_absolute_2024}. This demonstrates that the low-latency software infrastructure required for intelligent control is already maturing. However, the key challenge, as highlighted in recent reviews, is moving beyond general-purpose optimisers to more physics-aware algorithms \cite{yager2023autonomous}. This is essential because physical constraints inherent to \textit{operando} studies, such as path-dependency and hysteresis, invalidate simple, unconstrained optimisation strategies. This necessity for physics-informed event hunting has been successfully demonstrated by the ANDiE framework. To locate a magnetic phase transition, the ANDiE agent must be constrained to a monotonic heating path to avoid physically invalidating its results via hysteresis. This provides a direct foundation for the Heuristic \textit{Operando} framework, which extends this event-hunting concept from deterministic phase transitions to unpredictable, stochastic events \cite{mcdannald2022fly}.

%% file: Text/05_Conclusion.tex
\section*{Outlook and Conclusion}

In response to the recognised 3R challenge hindering \textit{operando} battery characterisation, this paper presented a critical analysis of the multi-modal electrochemical toolkit at the Rutherford Appleton Laboratory. This analysis confirmed that limitations related to Reliability, Representativeness, and Reproducibility are systemic and often tied to specialised hardware constraints. However, our analysis also highlighted a deeper methodological flaw: the inherent inefficiency and blindness of conventional, passive, pre-programmed experiments to the unpredictable, stochastic events critical to understanding battery degradation. To address these interconnected issues, we propose Heuristic \textit{Operando} experiments as a conceptual blueprint for the autonomous battery laboratories of the future: a methodological framework designed to transform passive observation into a rational, active search.

\subsection*{A New Definition of Experimental Efficiency}
This vision implicitly proposes a new definition of efficiency for experiments at large-scale scientific facilities. The current methodology often equates efficiency with data volume per unit time (throughput), a metric that directly contributes to the recognised data deluge. In contrast, Heuristic \textit{Operando} redefines efficiency as scientific insight per photon or neutron.

By adopting the Entropy-Scaled Measurement Efficiency ($E_{\mathrm{SME}}$) metric defined above, we shift the value proposition from volume to predictive power. For a battery researcher, the ultimate metric of efficiency is not collecting terabytes of cycling data, but deterministically identifying the specific failure mechanism in the first cycle that dictates the cell's end-of-life behaviour. By capturing the decisive moments that determine this trajectory, rather than the months of routine cycling between them, we maximise the impact of every second of beamtime. This represents a key philosophical shift, prioritising data quality and relevance over sheer quantity. The logical implication is a fundamental change in beamtime allocation strategies: instead of rewarding the volume of data acquired, success is measured by the deterministic capture of critical scientific phenomena, specifically the stochastic failure events targeted by this framework.

This redefinition is not just theoretical; it has quantifiable benefits. Recent proof-of-concept work is already validating this insight-per-neutron approach. An AI-steered workflow at the Spallation Neutron Source (SNS), for example, demonstrated the potential to save up to 29\% of neutron beamtime by halting data collection as soon as a pre-defined quality threshold was met \cite{mcdannald2022fly}. Even more striking, the ANDiE autonomous neutron diffraction system demonstrated a fivefold reduction in the measurements required to discover a magnetic phase transition. This work provides powerful validation for the core value of our framework: by making experiments smarter, we can not only capture new science but also dramatically improve the throughput and efficiency of our most valuable scientific instruments.

\subsection*{A Phased Implementation Roadmap}

Translating this framework from concept to reality involves a phased research program. The initial phase is an offline, computation-heavy task focused on building and validating the predictive models for the AI Pilot. This requires generating a vast library of HPC-generated digital twins to train and validate surrogate models (such as the GNNs or PINOs discussed previously) and rigorously testing them against historical data to bridge the sim-to-real gap. Concurrently, robust software interfaces must be engineered to couple these AI models to the live beamline environment. This integration requires developing low-latency data pipelines and an agent-based software layer capable of issuing high-level commands to existing instrument control systems, such as Bluesky or TANGO. Following commissioning—which may include a human-in-the-loop mode to build operator trust—the framework will be deployed for scientific discovery campaigns. This deployment initiates the final, most crucial step: a self-reinforcing feedback loop, where the unique, high-value data captured by the AI Pilot is fed back into the data platform to become the gold standard training set for the next, even more accurate, generation of AI Pilots.

Underpinning this loop is the requirement for a Findable, Accessible, Interoperable, and Reusable (FAIR) data platform. The relationship between the AI Pilot and this platform is dynamic and reciprocal. Initially, the AI Pilot requires curated FAIR data for its training. The AI-led experiments then generate new, high-value data that is uniquely contextualised, as the results are inherently tagged with the metadata explaining why they were acquired (e.g., scan triggered by precursor Z at coordinate Y). This richer dataset enriches the FAIR platform, becoming the superior training set for the next-generation AI pilot. This vision aligns with ongoing efforts at major facilities to enhance dataset FAIRness at the point of acquisition. For instance, recent work integrating ML-based online analysis into closed-loop synchrotron experiments has focused on architectures that embed real-time analysis results directly into facility data streams (e.g., NeXus HDF5 files) and electronic logbooks \cite{pithan2023closing}. This immediate integration of processed results and metadata is the crucial mechanism for realising this feedback loop, ensuring that AI-driven experiments automatically generate richer, more reusable datasets.

\subsection*{Governance and Trusted Autonomous Science}

The deeper vision of this framework extends beyond automation to the establishment of trusted autonomous science. This objective is a central pillar of the methodology; the task is not merely building the AI, but validating it, quantifying its uncertainty, and actively mitigating the confirmation biases that can arise in any closed-loop system. This explicit focus on governance provides the fundamental solution to the 3R challenge. While the intelligent search tackles the symptom (the statistical missing of stochastic events), this governance framework addresses the root cause: the reliance on unverifiable, operator-dependent decision-making.

Developing effective Human-in-the-Loop (HITL) frameworks, where expert intuition is integrated with AI decision-making, is a crucial step in this direction. As demonstrated by Adams et al. \cite{adams2024human} for autonomous phase mapping, such HITL approaches, coupled with interpretable visualisations, are essential for building user trust and ensuring that autonomous systems enhance, rather than obscure, scientific understanding. Establishing similar mechanisms for interpretability and expert oversight will be paramount for validating the predictions and actions of the Heuristic \textit{Operando} AI pilot.

In conclusion, the Heuristic \textit{Operando} Framework establishes a necessary reference architecture for next-generation research facilities. This perspective outlines a critical methodological shift, moving beyond passive, pre-programmed experiments to a unified, autonomous research capability. By synthesising HPC-driven Digital Twins, automated workflows, and real-time data infrastructures, the framework demonstrates that these technologies are not merely operational conveniences, but scientific necessities for capturing stochastic failure mechanisms. Consequently, this work provides a rigorous justification for the broad investment in infrastructure required to support trusted autonomous science. Ultimately, this transition, from passive data collection to an active, intelligent search, offers the most systematic pathway to accelerate discovery and resolve the critical materials challenges ahead.

%% file: Text/06_Methods.tex
\section*{Supplementary Information}

\subsubsection*{X-ray techniques [Veronica Celorrio, Isabel Antony]}
Synchrotron facilities, such as Diamond Light Source, provide high-brilliance, high-coherence X-ray beams with tunable photon energy \cite{drnec}. The available energy range spans from soft X-rays (typically 0.1--2 keV) to hard X-rays with shorter wavelengths and higher energies \cite{lin}. Incident beam energy is a critical parameter; alongside flux (intensity), it dictates the material's absorption cross-section. Typically, lower energies result in higher absorption and thus more severe radiation damage (dose) for a given flux \cite{christensen2023beam}. While hard X-rays are deeply penetrating and interact minimally with the bulk structure compared to soft X-rays, they carry significantly higher energy per photon. Consequently, secondary electron cascades resulting from absorption events can be severe. This radiation damage becomes a critical limiting factor in time-resolved \textit{operando} experiments, where the cumulative dose effectively concentrates in a single sample volume over the scan duration \cite{bras}. Therefore, carefully modulating the beam using filters is essential to maintain result accuracy and reliability. Despite these challenges, X-rays remain a crucial tool for studying battery materials, where employing a wide range of energies allows for complementary structural and dynamic information across different probing depths.
\par \vskip 0.25cm \indent X-ray techniques fall into three broad categories: scattering, spectroscopy, and imaging. The custom cell design for a particular instrument depends directly on the geometry of the specific analysis technique adopted on the beamline.
\par \vskip 0.25cm \indent X-ray diffraction (XRD) is an elastic scattering technique utilized when the X-ray wavelength is comparable to interatomic distances in the crystalline sample. The resulting scattered waves are measured by a ring of detectors as a function of scattering angle. This setup is exemplified by the I11 beamline at Diamond Light Source \cite{thompson}. Consequently, the \textit{operando} AMPIX cell used on I11 follows a similar geometry to the POLARIS cell, as both are customised for scattering setups. XRD determines structural information including lattice parameters, average crystal size, strain, and crystalline phase. \textit{In operando} XRD is frequently adopted to inspect structural changes in battery components during electrochemical processes.
\par \vskip 0.25cm \indent Complementarily, X-ray Pair Distribution Function (XPDF) analysis offers insights into local structure. Also known as total scattering analysis, this method subjects both Bragg peaks and diffuse scattering components (the total scattering pattern) to a Fourier transform to obtain the PDF. The PDF, $G(r)$, gives the probability of finding an atom at a distance $r$ from another given atom. Because it probes materials regardless of long-range order, it is especially useful for characterising nanocrystalline, fluid, and amorphous substances. The I15-1 beamline at Diamond is dedicated to this technique, utilising the DRIX cell designed specifically for PDF measurements. The DRIX design is optimised to produce an insignificant, consistent, and reproducible background—a requirement crucial for accuracy and reliability in total scattering.
\par \vskip 0.25cm \indent Spectroscopic X-ray techniques include X-ray Absorption Spectroscopy (XAS) and X-ray Photoelectron Spectroscopy (XPS). Akin to PDF, XAS applies to materials with or without long-range order. Each core electron excited by the incident beam results in an absorption edge, the spectrum of which yields information on oxidation state and bond length. XAS data is typically split into two regions: X-ray Near Edge Structure (XANES) and Extended X-ray Absorption Fine Structure (EXAFS). The edge energy position in the XANES region identifies the oxidation state of the atom of interest through comparison with standards. Additionally, pre-edge features describe local geometry, while the overall spectral shape reflects the electronic structure around the probed atom \cite{lin}. The EXAFS spectrum allows for qualitative analysis of the average local structure in electrode materials lacking long-range order. B18 at Diamond is a general-purpose XAS beamline where extensive facilities allow users to probe electrocatalyst electrochemistry \textit{in operando}. Users have access to three custom cells: a flow cell, a static cell, and a cell optimised for studying gas diffusion electrocatalyst electrodes.
\par \vskip 0.25cm \indent XPS studies photoelectrons emitted from a sample after excitation by incident X-ray photons. Being surface-sensitive (probing only a few nanometres deep), it yields elemental and chemical composition data for surfaces, complementing the bulk sensitivity of XAS. The spectro-electrochemical flow cell available on the B07 beamline at Diamond is optimised for soft X-ray spectroscopy (both XAS and XPS). Because B07 operates at lower energies (soft X-rays), some experiments require vacuum conditions; the cell is therefore designed to maintain stack pressure within a vacuum environment.
\par \vskip 0.25cm \indent Finally, X-ray imaging techniques leverage photoelectric absorption. This interaction varies with the atomic number $Z$ of the absorber, scaling approximately as $Z^4$. To reveal internal structure non-destructively, two-dimensional shadow pictures (radiographs) are taken from many angles (usually 180 or 360 degrees) and reconstructed computationally into a three-dimensional object (tomography). A variety of sample environments are used across beamlines and lab-based machines, ranging from commercial cells and \textit{ex situ} samples to bespoke lab-built designs such as coin or pouch cells \cite{xu2025operando} (see section \textit{Coin and pouch cells}).
\par
\subsubsection*{Neutron techniques [Gabriel Perez, Scott Young]}
    Neutron analysis techniques produce complementary information to X-ray techniques due to their sensitivity to light elements, such as lithium, oxygen and hydrogen. They are also able to discriminate among elements with similar electronic configurations such as the 3d-block transition metals. These properties make them uniquely suited to battery experiments; for example, neutrons are used to probe lithium distribution and occupancy in an \textit{in operando} cell as well as determine any degree of cation intermixing in the structure. Furthermore, different isotopes have significantly different scattering lengths. Hence, samples can be isotopically enriched to improve the signal-to-noise ratio of the data and enhance contrast between components of interest. For example, the  incoherent scattering from hydrogen-containing samples can be substantially reduced by deuterating them. Another common application of isotopic enrichment in battery materials is the use of \(^{6/7}\)Li substitution, since \(^6\)Li has a high neutron absorption cross-section, whereas \(^7\)Li will minimise absorption \cite{saito}. In this way, sections of samples may be labelled by an isotope to highlight their contribution to the scattering spectrum or improve the accuracy of the data analysis. Neutrons are also highly penetrating, and only weakly interact with the sample they pass through. Thus, radiation damage is not an common issue when utilising neutron analysis techniques.
\par \vskip 0.25cm \indent
    Akin to X-ray techniques, neutron methods can be divided into spectroscopy, scattering and imaging. As before, the instrument set-up and instrument geometry will influence the corresponding cell design.
\par \vskip 0.25cm \indent
    The principle of neutron diffraction is identical to X-ray diffraction, but yields complementary information. It is one of the most commonly used methods to characterise materials during fundamental research. For battery material research, powder diffraction is typically preferred over analysing single crystals, which are difficult to make and not representative of commercial battery materials. This method is offered by several different instruments across the ISIS Neutron and Muon Source site. Most notably, POLARIS and NIMROD have hosted several cell experiments. Recently, diffraction detectors have been added onto IMAT. Although they are not yet in operation at the time of this report, they will be able to measure neutron diffraction data in tandem with the existing detectors for future experiments, albeit at a limited Q-range. It is interesting to note that since the geometry and beam spot sizes of neutron diffraction, small angle neutron scattering (SANS), and muon spectroscopy instruments are comparable, their respective custom cells can often be used interchangeably, or modified slightly to suit the other techniques. For neutron diffraction, the most prominent and versatile example of a custom cell available at RAL is the POLARIS cell.
\par \vskip 0.25cm \indent
    SANS is a scattering technique where elastically-scattered neutron radiation is measured and detected at small scattering angles. It has predominately been used in battery material analysis to probe dynamic information about pore contents in the electrodes, as well as observe the formation of the solid electrolyte interface (SEI) \cite{perez}. There are several instruments at ISIS that are able to employ this technique. NIMROD is optimised for PDF and SANS measurements; it specialises in multi-scale analysis of disordered materials \cite{headen}. In particular, it has also been utilised to perform neutron diffraction on electrolytic samples. This has allowed relations to be drawn between the macroscopic behaviour of electrolytes, and the microscopic properties of hydrogen bonds which act as defects within such structures. Users have observed how the number and stability of hydrogen bonds influence the diffusion, viscosity and conductivity of electrolytes - critical properties that can be manipulated to boost the performance of batteries \cite{busch}. Although a cell has been designed for NIMROD, its geometry is not optimised for battery research; efforts are instead being directed to focus on the creation and development of a pouch cell holder for future experiments. Furthermore, the BAM cell has been adapted for the SANS technique and used at SINQ \cite{reynolds}. The BAM cell has also been adapted to be used for neutron powder diffraction on Polaris and the success of these experiments reinforces the idea of commutability between cells when used for techniques with similar geometries, as previously discussed.
\par \vskip 0.25cm \indent
    Imaging allows for features to be probed in real space in a variety of techniques. The IMAT instrument of ISIS is able to carry out radiography, tomography and Bragg edge imaging. Radiography involves producing a shadow by placing the sample in the path of the neutron beam and taking an image of the transmitted beam. Tomography elevates this concept by combining multiple radiographs taken at different angles around the sample to form a 3D image. Through these imaging methods, the evolution of defects in the sample can be seen. Specifically in the context of batteries, this enables the study of the formation of new lithiated and delithiated phases as a function of charge/discharge and the growth of dendrites, in addition to monitoring gas bubble formation and lithium distribution across electrodes \cite{ziesche2022neutron}. Furthermore, formation of gas bubbles and lithium distribution across electrodes can also be studied. Bragg edge imaging is based on the drop in transmitted intensity due to Bragg scattering away from the incident beam direction for a particular set of crystallographic planes that satisfy the Bragg condition. When the Bragg condition is no longer satisfied (at $\lambda > 2d$), there is a sharp rise in transmitted intensity, giving rise to characteristic Bragg edges. Consequently, this imaging technique corresponds directly to the crystalline structure of the material, providing spatially resolved information on texture, phase distribution, and lattice strain \cite{ramadhan}. Currently, no specialised cells have been designed for the IMAT instrument, but users often bring their own cells and sample environments to beamtimes.
\par
\subsubsection*{Muon techniques [Peter Baker, Tom Wood]}
    Muon spectroscopy ($\mu^+$SR) is a bulk technique often used to probe the diffusion of ions, particularly lithium and sodium ions. Other ions may also be investigated with this technique, as long as they have a relatively sizeable nuclear magnetic moment and substantial abundance. Table~\ref{fig:muon_ions} below, adapted from McClelland et al.'s paper, summarises the suitability of ions to the muon spectroscopy technique \cite{mcclelland_diffusion}. The technique relies on implanting muons into the component of interest, and detecting the positron, or less commonly, electron for ($\mu^-$SR), decay product that is emitted in the direction of the muon spin at the moment of decay \cite{blundell}. Hence, the BAM cell, created for the EMU instrument at ISIS and optimised for muon spectroscopy, can be adjusted to a range of thicknesses so that the incident muons stop in the component to be studied.
\par
\begin{table}[h]
    \begin{center}
    \begin{tabular}{|c|c|c|c|} 
        \hline
        \textbf{Nucleus} & \textbf{Moment} & \textbf{Abundance} (\(\%\)) & \textbf{Chance of success}\\
        \hline
        \(^1\)H & +4.84 & 99.9885 & Have to separate \(\mu\)+ and H+ motion\\ 
        \hline
        \(^7\)Li & +4.20 & 92.41 & Excellent and well-studied\\ 
        \hline
        \(^{19}\)F & +2.63 & 100 & Difficulties due to F-\(\mu\) bond formation\\ 
        \hline
        \(^{23}\)Na & +2.86 & 100 & Excellent; some work has been done\\ 
        \hline
        \(^{25}\)Mg & -0.86 & 10.13 & Works but with a small signal\\ 
        \hline
        \(^{27}\)Al & +3.64 & 100 & Promising\\
        \hline
        \(^{39}\)K & +0.39 & 93.08 & Works but with slower relaxation rate\\
        \hline
        \(^{43}\)Ca & -1.49 & 0.135 & Poor unless enriched\\
        \hline
        \(^{51}\)V & +5.15 & 99.76 & Promising\\
        \hline
        \(^{67}\)Zn & +0.875 & 4.1 & Poor unless enriched\\
        \hline
        \(^{127}\)I & +2.81 & 100 & Excellent\\
        \hline
    \end{tabular}
    \end{center}
    \caption{Table showing the nuclear magnetic moment, abundance and chance of success of ions for muon spectroscopy, from McClelland's et al.'s paper \cite{mcclelland_diffusion}}
    \label{fig:muon_ions}
\end{table}

\newpage
\subsection*{POLARIS cell [Gabriel Perez]}
\subsubsection*{Overview of the POLARIS cell}
    The POLARIS instrument, situated in Target Station 1 (TS1) of the ISIS Neutron and Muon Source facility, is a high-intensity, medium-resolution neutron powder diffractometer \cite{smith}. It has been the site of extensive battery research, culminating in several publications since starting its operations. The POLARIS cell is the custom cell for this instrument, developed for \textit{in operando} studies on electrode materials. It has since allowed for observation of the structural processes that occur owing to electrochemical reactions \cite{biendicho}.
    \subsubsection*{Design of the POLARIS cell}
    The cell consists of circular components, which are stacked together, resulting in a structure reminiscent of the typical coin cell geometry. The area exposed to the beam measures 4cm\(\times\)1.5cm (height\(\times\)width), which is reflected in the aperture dimensions of the POLARIS cell \cite{smith}.
\par
\begin{figure}[h]
    \centering
    \includegraphics[scale=0.25]{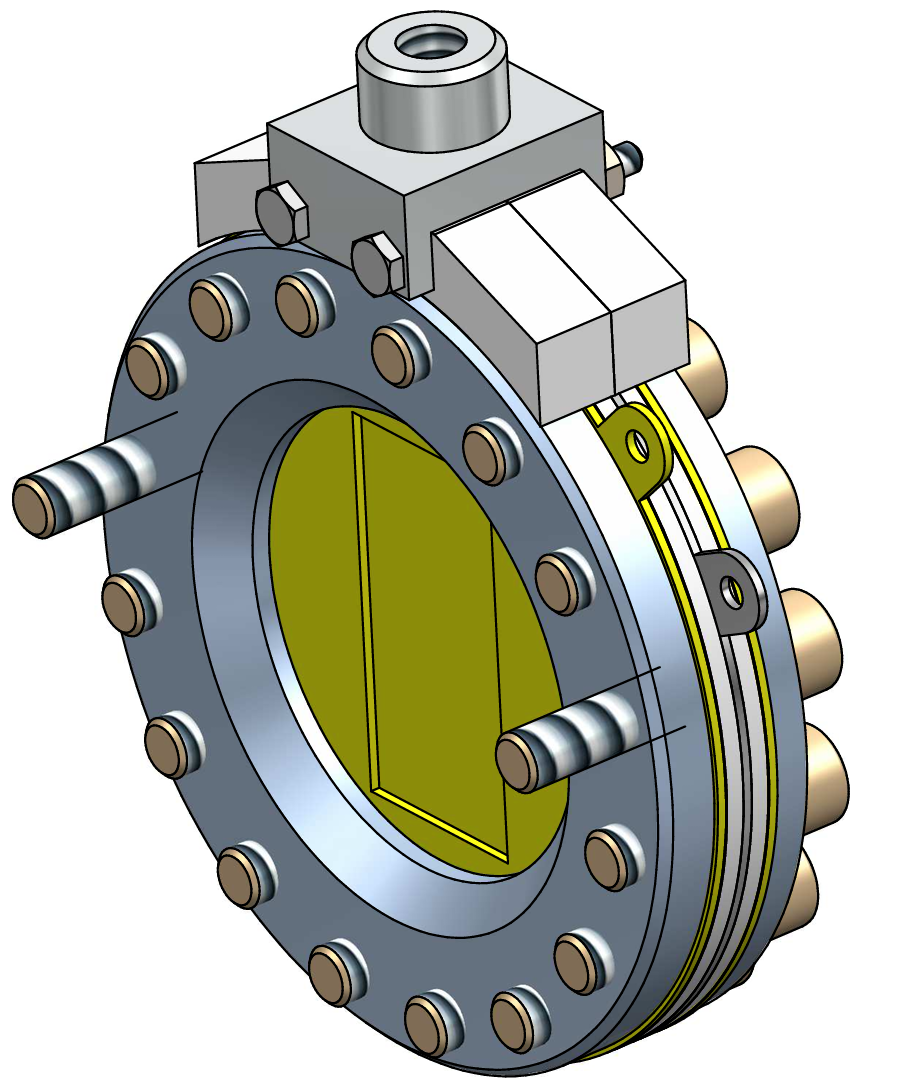}
    \caption{CAD drawing of POLARIS cell}
    \label{fig:polaris_cad}
\end{figure}

 \vskip 0.25cm \indent
    The cell is highly modular, and can be adapted to form full cells or more simply, half-cells. The typical architecture of a POLARIS battery is described in this paragraph, although cell configurations can be varied for different purposes. At the centre of the cell, there is a 2mm thick separator, which is made of a chemically inert plastic that prevents it from reacting with the electrolyte. The separator is also electrically insulating and separates the two opposite electrochemically charged sides of the cell. On each side of the separator, there is a 0.2 mm vanadium current collector on which the electrodes will be placed (typically coated or cast). The separator and adjacent current collectors are sandwiched by two thick metal window plates with a protuberance for connecting the cell to the potentiostat. The whole assembly is then clamped by two thick plates and the entire cell is fixed in place with sixteen polyether ether ketone (PEEK) screws, ensuring a hermetic seal with the aid of O-rings between the separator and the two adjacent current collectors. The separator and metal windows used as terminals have a 2x4 cm aperture to allow the beam pass through them and avoid scattering from those components. Typically, the electrodes will be coated with this geometry as well and their placement will be aligned with the aperture to minimise the electrochemically and neutronically inactive volume of the cell. The aperture of the separator will be filled with a glass fibre sheet, which will be wet by electrolyte. The whole assembly is shown in Figure~\ref{fig:polaris_cad}, orientated as if the beam were pointing into the page. A boron nitride shield is placed on the front of the cell after assembly to absorb any stray neutrons that are not a part of the collimated beam, but also to absorb the scattered neutrons from the components of the cell. A mount is also added to the top of the cell to attach it to the stick that will go in the instrument.
\par
\indent
    The windows also act as the current collectors and can be made from four different elements: copper, aluminium, vanadium, and nickel. The choice of material allows the user to have some control over where in the diffraction patterns the peak from the current collectors and what their contribution to the background will be. For instance, to facilitate data processing, it is recommended to use a material whose diffraction pattern minimises overlap with the sample data. Regarding the cell assembly it is also worth mentioning that the cell can be have a stacked architecture with two repeating electrochemically active units to maximise the signal of interest.
\par
\subsubsection*{Experimental methodology to use the POLARIS cell}
    Prior to the beamtime, it is strongly recommended that users spend time optimising their system of interest in the cell. A good balance between electrochemical performance and data quality must be achieved. The first important decision is the window material. For example, although vanadium is the most transparent to neutrons and would thus permit the most transmission of signal, it is the least preferable from an electrochemical point of view. The second consideration is the thickness of the electrode of interest. Although a thicker electrode would yield a better data quality, it would also worsen the electrochemical performance of the cell. To mitigate this, users should attempt to operate the cell with varying electrode thicknesses, starting with replicating the processing conditions of their coin cell electrode to obtain a similar thickness and incrementing until an acceptable electrochemistry is reached. The same principle applies to the cathode mix with binder and carbon. The larger mass loading in the Polaris cell can require a higher amount of conductive carbon and binder for good electrochemical performance and cathode robustness.
\par \vskip 0.25cm \indent
    Neutron diffraction is the technique on POLARIS; as such, samples containing hydrogen are often deuterated to reduce the incoherent contribution to the background. To determine the level of deuteration needed, it is recommended to first take readings of a non-deuterated cell. Based on the background signal given, the deuteration can be adjusted accordingly. Ideally, a fully deuterated electrolyte should be used, but in cases when this is a scarce resource, the electrolyte for the experiment could be partially deuterated up to acceptable levels to not impact the background of the cell substantially.
\par \vskip 0.25cm \indent
    It is highly recommended to have a database of the individual components of the cell prior to the experiment to aid determination of contributors to the diffraction pattern of the cell. At the very least, a measurement of the pristine powder is strongly recommended to have. Some of the common materials and components of the cell have already been measured and are available to the community. The user can enquire with their local contact for access to these. At the time of writing, two potentiostats are available for cell cycling. The obtained results can be compared to the benchmark signal for analysis.
\par
\subsubsection*{Applications of the POLARIS cell}
    Compared with a conventional coin cell, the POLARIS cell has a substantially higher mass loading. As a result, ion diffusion is more limited, which significantly reduces electrochemical performance; the POLARIS cell operates at a level that is not representative of commercial applications. Thus, the POLARIS cell has a low Technology Readiness Level (TRL) and is best suited to fundamental studies of new electrode materials. Results obtained are not directly transferable to conventional coin cells.
\par \vskip 0.25cm \indent
    Neutron techniques can be sensitive to light elements, namely lithium and hydrogen, and elements with similar electronic configurations than their X-ray counterparts \cite{biendicho}. Thus, \textit{in operando} neutron diffraction experiments are often used to observe the structural processes and phase changes that involve lithium, sodium, and various 3d transition metals that occur at the electrodes during electrochemical cycling \cite{srinivasan}. \textit{In operando} neutron diffraction experiments can also determine important crystal structural parameters of the sample through Rietveld analysis \cite{taminato}. As stated by Perez et al. in their review paper, other information that can be gained through neutron diffraction includes magnetic structure, M-O bond lengths, cation oxidation state and distribution of cations in TM cathodes \cite{perez}.
\par \vskip 0.25cm \indent
    All cells that can be used on POLARIS may also be utilised on GEM and WISH. Furthermore, owing to their similar experimental geometries, cells which are custom built for neutron diffraction techniques may be adjusted for muon spectroscopy and other neutron techniques. However, due to the smaller beam size at these instruments, the POLARIS cell is not optimised for these applications. Moreover, the current collector material may need to be substituted for one that is more suitable to the analysis technique. For (\(\mu\)SR), this is typically stainless steel, which is both transparent to muons and non-magnetic. For SANS, current collectors are typically aluminium, as it is relatively transparent to neutrons, but also does not contribute to the signal on a nanometre scale. As SANS is sensitive in the nanometre region, this latter property is essential for good data quality.
\par
\subsubsection*{Known limitations of POLARIS cell}\label{sec:polaris_limitations}
    This cell design brings about some limitations that may introduce challenges to users when carrying out their experiment. Firstly, the stack pressure of the POLARIS cell is not standardised, and there are no current methods to take a quantitative measurement of this pressure. Pressure is concentrated around the PEEK screws, and inconsistent across the face of the cell. Since POLARIS will typically operate under vacuum conditions, these inconsistencies may lead to unreliable and irreproducible results, as a higher stack pressure could cause better electrode contact with the current collectors. To mitigate this, the PEEK screws are tightened as much as possible to ensure there is full contact.
\par \vskip 0.25cm \indent
    Furthermore, due to the large area of the cell, the electrolyte will eventually settle at the bottom. Consequently, the top of the cell will become electrochemically inert, culminating in heterogeneous cycling, which causes unreliability and irreproducibility in results. Users are encouraged to bring in their cell materials in case new cells need to be synthesised on-site in case of cell failure.
\par \vskip 0.25cm \indent
    There are also limitations to the types of experiments that may be carried out with the POLARIS cell. Currently, a custom heating and cooling environment is being developed for the cell, which will enable experiments to be performed at temperatures between \(-20 ^\circ\)C and \(80 ^\circ\)C. This will facilitate solid state battery studies, an area previously unexplored with POLARIS cells.
\par

\newpage
\subsection*{Electrochemical cell for combined SANS and Total-Scattering on NIMROD [Tom Headen, Ali Mortazavi]}
\subsubsection*{Overview of NIMROD and dedicated electrochemical cell}
    The NIMROD instrument is a diffractometer designed to study disordered materials over a very wide $Q$ range ($0.02 < Q < 50 \text{\r{A}}$). A flexible electrochemical cell was designed for the instrument, in collaboration with QMUL and UCL. This allows it to be used in a variety of electrochemical experiments including of electro-catalysis, supercapacitors and batteries. 
\par
\subsubsection*{Design of the NIMROD cell}
\begin{figure}[h]
    \centering
    \includegraphics[scale = 1]{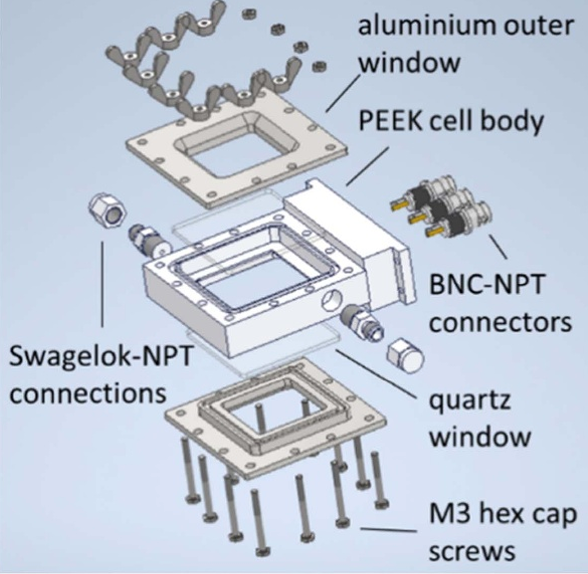}
    \caption{Schematic of the NIMROD cell, taken from Figure 15 of Shah et al.'s paper \cite{shah}}
    \label{fig:NIMROD_cell}
\end{figure}

    As seen from Figure~\ref{fig:NIMROD_cell}, the main cell body is formed from PEEK. The NIMROD cells notably employ quartz windows, as opposed to the materials used within POLARIS cells. Quartz does not contribute any Bragg peaks to the scattering data, which could overwhelm the underlying disordered diffuse scattering. Furthermore, since quartz is a glass, it has no long-range order, and thus is suited towards SANS experiments. The windows are larger than the 30\(\times\)30mm maximum beam size on the NIMROD instrument. The cell has a 1mm path length to allow the use of hydrogenated electrolytes which would otherwise attenuate the beam too much. Hydrogentated electrolytes may be useful to provide maximum SANS contrast to a carbon electrode, or to be run in combination with deuterated electrolytes to exploit H/D isotopic substitution, or simply because deuterated electrolytes are not available.

\subsubsection*{Experimental process of NIMROD cell}
   The cell is flexible, in that any electrochemical setup can be built by removing the front loading windows. This can then be sealed with O-rings (a requirement due to the vacuum in the instrument) and electrolyte added through the side ports to flood the cell. It is also possible to continuously flow electrolyte through these ports and through the cell during an experiment. 

   \subsubsection*{Applications of the NIMROD cell}
    NIMROD can employ total scattering and H/D substitution to allow the study of liquid structure in unprecedented detail, and has been used to study important new electrolytes such as water-in-salt systems through \emph{ex situ} studies \cite{groves2025lithium}. This cell could therefore be used to study electrolyte systems under \emph{operando} conditions using this cell. Furthermore the wide $Q$-range of NIMROD is particularly suited to the study of liquids confined in micro and mesoporous materials and so would be a unique way to study processes in hard carbon battery anodes and porous carbons used in supercapacitors. Lastly, outside of energy materials, the cell has can be used to study solution structure of novel oxidation states of solutes and for other electrochemical systems such as in electrocatalysis. 

\subsubsection*{Limitations of the NIMROD cell}
    Due the the high flexibility of the design it is not optimized for the study of batteries, for example it can be quite difficult to assemble a battery inside inert environments, particularly in a glovebox. Further designs are currently being built with more specialised designs focused on electrocatalysis and for supercapacitor studies. 

\newpage
\subsection*{Battery Analysis by Muon (BAM) cell [Peter Baker]}
\subsubsection*{Overview of the BAM cell}
    Originally designed for the EMU instrument at ISIS, the BAM cell is the first electrochemical cell customised for \textit{operando} experiments using $\mu$SR \cite{mcclelland} on any instrument. Variations have since been made to adjust the cell to other techniques, including a version modified for SANS \cite{reynolds}. 
\par
\subsubsection*{Design of the BAM cell}
    For a full discussion of the BAM cell structure, readers are directed to McClelland et al.'s thorough explanation in Chapter 6 \cite{mcclelland}.
\par \vskip 0.25cm \indent
    Similar to the POLARIS cell, the BAM cell is formed through layering components to form a battery with a coin cell adjacent geometry. The centre of the cell consists of a fluorosilicone gasket, with an circular aperture of diameter 18mm to hold the battery. This is held between two stainless steel windows; they in turn are sandwiched between two stainless steel holders. The outermost layers of the front and back consist of a silver mask which collimates the beam and prevents any stray muons from reaching the sample. A total of eight PEEK screws are used to lock the layers together and form a seal that remains hermetic under high vacuum.
\par \vskip 0.25cm \indent
    The windows hold a dual purpose as current collectors. Their thickness (from 50-100\(\mu\)m) controls the depth of muon penetration, permitting different parts of the cell to be studied. A thicker window will be more rigid - this helps to provide better contact with the corresponding electrode, optimising the cell's electrochemical performance. Although stainless steel is typically used due to its non-magnetic property, in variations of the BAM cell other materials may be preferred, making it more similar to the POLARIS cell but with a smaller active area.
\par \vskip 0.25cm \indent
    Gaskets of varying thicknesses are available (from 0.4-1.5mm), and can be stacked to form cells with different thicknesses. This dimension is crucial because it must be sufficiently thick to maintain adequate electrical contact between the current collectors and the electrode, as well as ensure structural integrity. However, this thickness must not hinder the electrochemical performance of the cell.
\par
\begin{figure}[h]
    \centering
    \includegraphics[scale = 0.25]{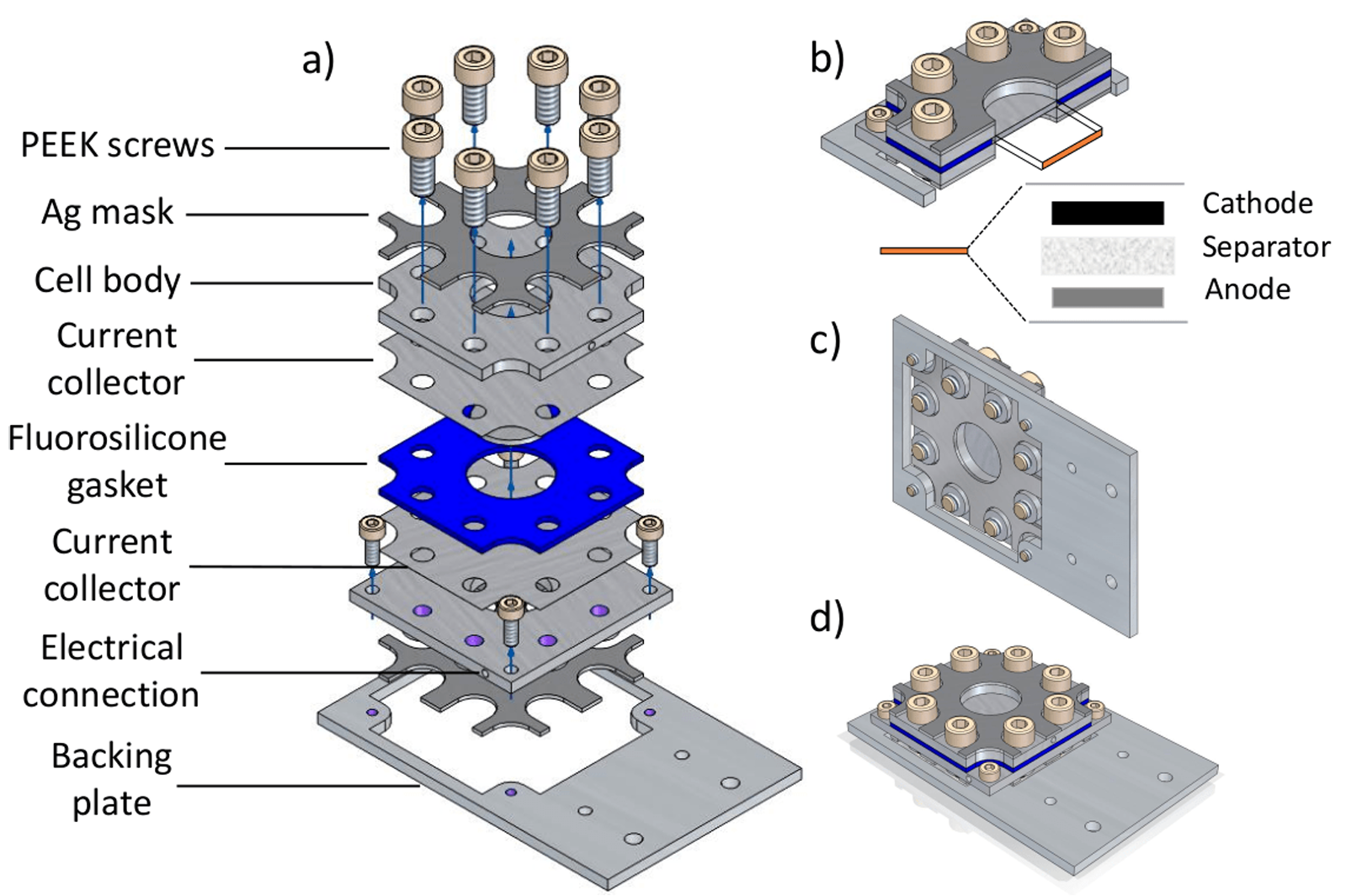}
    \caption{Diagram showing various views of the BAM cell, taken from Figure 6.1 of McClelland et al.'s paper \cite{mcclelland}}
    \label{fig:bam_cad}
\end{figure}

 \vskip 0.25cm \indent
    This design ensures that the BAM cell is fully reversible, and can be orientated such that either electrode or an electrolyte may be studied. It is shown in Figure~\ref{fig:bam_cad} in a series of different views, including a cross-section view (b).
\par
\subsubsection*{Experimental process using the BAM cell}
    As mentioned above, due to the modular design of the cell, the window thicknesses can be varied. When this is controlled in conjunction with the beam degraders to change the muon implantation depth, different parts of the cell can be studied.
\par
\subsubsection*{Applications of the BAM cell}
    The BAM cell is optimised for muon spectroscopy on EMU, which can be carried out using both positive and negative muons. Positive muon experiments have been carried out by McClelland et al. to inspect the motion of lithium ions as the \textit{in operando} cell undergoes electrochemical reactions, and thus, the degradation mechanisms that occur over multiple cycles. Although no such experiments have been performed thus far, McClelland et al. also remark on the possibility of using negative muons for elemental analysis \cite{mcclelland_diffusion}.
\par \vskip 0.25cm \indent
    As previously discussed in Table \ref{fig:muon_ions}, muon spectroscopy is only sensitive to the motion of ions with nuclear magnetic moments, as for NMR, but it is less sensitive to the presence of paramagnetic ions. Ions that have been studied include Na$^+$, K$^+$ and I$^-$ \cite{wright} \cite{matsubara} \cite{ferdani}. However, there is some potential in conducting experiments with magnesium, aluminium and vanadium.
\par \vskip 0.25cm \indent
    The design of the BAM cell is modular, useable on other muon instruments, and adaptable to other techniques, such as SANS. In this application, the windows were constructed from aluminium instead of stainless steel due to their transparency to neutrons. However, the BAM cell along with any of its variations is of a low TRL, and any results produced may not be representative of commercial cells.
\par \vskip 0.25cm \indent
    Additionally, both liquid and solid electrolyte cell configurations have been studied using the BAM cell, with the latter carried out above room temperature. Wider temperature dependence studies have not yet been carried out but the cell design should be able to operate over a similar window to commercial cells.

\subsubsection*{Known limitations of the BAM cell}
\par \vskip 0.25cm \indent
    When using the cell for muon spectroscopy, the sample and cell materials cannot be magnetic at the operating temperature and nuclear magnetic moments in the cell materials are avoided to give an easier to subtract background signal, mitigated by masking the area around the window using silver. For elemental analysis experiments using negative muons, any elements of interest in the sample should not also be used to construct the sample holder.
\par \vskip 0.25cm \indent  
    It was found that for high cathode mass loading configurations the BAM cell produced cycling performance representative of a low mass loading Swagelok cell at C/20 and C/10 \cite{mcclelland}. However, owing to an inevitable increase in internal impedance as a result of a thicker cathode it was not able to manage the high current rate required for C/5. Furthermore, the high mass loading of the BAM cell is likely to result in more capacity fade. This will be accentuated under higher cycling rate conditions. Hence, users need to find a suitable balance between good electrochemical performance and quality of data obtained.
\par \vskip 0.25cm \indent
    To conduct studies above or below room temperature, the chamber in which the BAM cell is held must be evacuated. In this case, thicker current collectors are needed to avoid them buckling outwards, which reduces the stack pressure and electrical contact. Similar to the POLARIS cell, the stack pressure is not currently set in a reproducible or quantified manner. While control is possible in principle, the reliance on manual torque without integrated sensors introduces a variability that impacts experimental consistency.
\par

\newpage
\subsection*{DRIX cell [Dan Irving]}
\subsubsection*{Overview of the DRIX cell}
    In order to carry out \textit{in situ} PDF measurements, constant and reproducible backgrounds are required, with amorphous materials being preferred. To achieve this, the sample holder must not introduce Bragg peaks, and the peripheral cell components must work to maintain homogeneous electrochemical cycling. The Diamond Radial \textit{In Situ} X-ray (DRIX) cell was developed with these aims in mind. It is available for the I15-1 user community, but has also been used on DIAD. In both cases, the beam dimensions fall on the micron scale, allowing finer resolution, enhancing the clarity when local scans of the sample are undertaken.
\par
\subsubsection*{Design of the DRIX cell}
    The passage below provides a succinct overview of the DRIX cell, highlighting the most important features. A full analysis can be found in Diaz-Lopez et al.'s paper \cite{diaz-lopez}.
\par \vskip 0.25cm \indent
    Instead of the axial approach of the BAM and POLARIS cells, the DRIX cell adopts a radial geometry. This isolation of components allows the beam to penetrate a larger volume of the layer of interest, and also reduces the number of obstacles in the beam path. Thus, the proportion of signal obtained from the sample material is maximised.
\par
\begin{figure}[h]
    \centering
    \includegraphics[scale = 0.25]{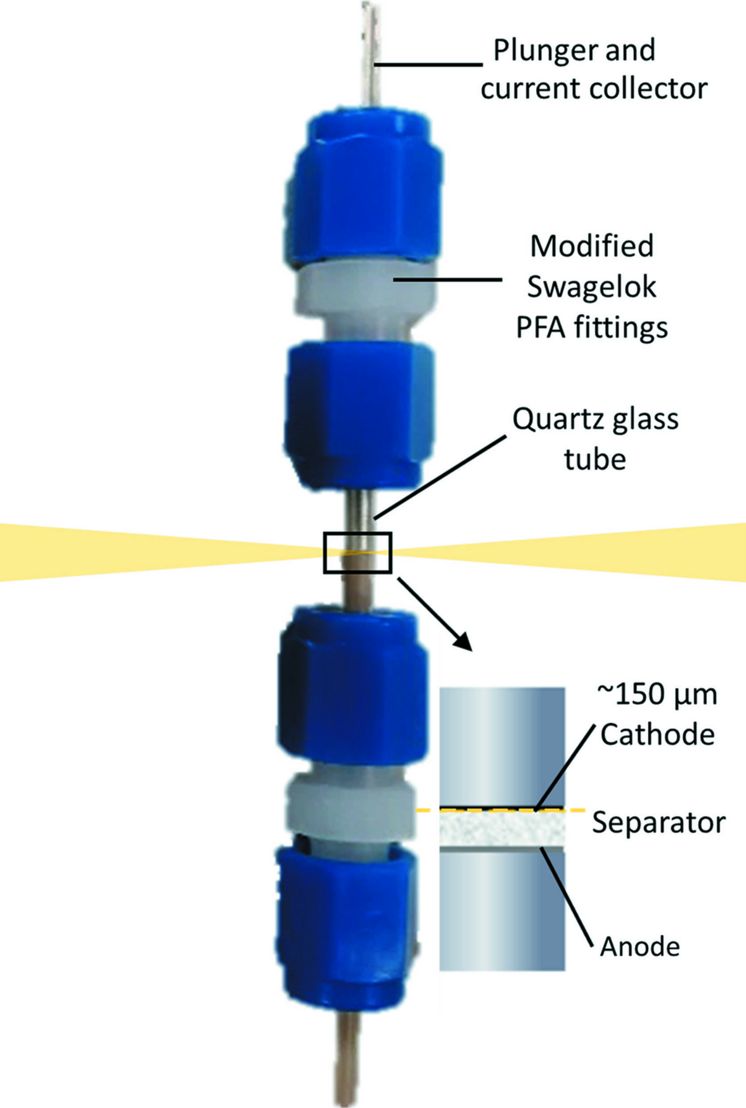}
    \caption{Photograph of the assembled DRIX cell, taken from Figure 1 of Diaz-Lopez et al.'s paper \cite{diaz-lopez}}
    \label{fig:drix_photo}
\end{figure}

 \vskip 0.25cm \indent
    The current generation of DRIX cell is simply comprised of four parts: plastic unions, glass capillaries, current collector rods, and ferrules. Notably, unlike other specialised cells reviewed here, the DRIX design relies primarily on modified off-the-shelf commercial components (e.g., standard Swagelok PFA fittings ), rather than bespoke machined parts. This reliance on standardised hardware significantly lowers the barrier to fabrication and enhances inter-facility reproducibility. Additionally, small springs are sometimes fitted into the assembly to ensure that the stack pressure is sufficient for adequate electrochemical performance. This assembly is shown in Figure~\ref{fig:drix_photo}, along with the position of the X-ray beam.
\par
\indent
    For the cell body, a thin-walled fused-quartz glass tube of 3.18mm (outer diameter) and 2.9mm (inner diameter) is employed, plugged at both ends by a Swagelok union and ferrule to maintain a hermetic seal. It also acts as the sample holder, containing the electrodes and electrolyte during the experiment. In the first iteration of the DRIX cell design, the cell body was composed only of two Swagelok unions. However, as their deformation and mechanical instability resulted in an irreproducible background, the switch to glass was made. Moreover, glass has insulating and optical transparency properties which simplifies cell design and beam alignment respectively.
\par \vskip 0.25cm \indent
    The Swagelok unions attach the cell body to current collector rods, which are usually formed from vitreous carbon due to its low absorption profile and good conductivity. Aluminium and stainless steel are also sometimes used, despite the high absorption profile of the latter material.
\par
\subsubsection*{Experimental process using the DRIX cell}
    A stainless steel multi-cell holder is available to probe several cells at a time. The beam sweeps across the vertical cells in the direction perpendicular to its propagation. Eight batteries can be cycled in parallel using the potentiostat available. The holder position is calibrated using a moving frame that can be adjusted in the x and y directions with a micron accuracy.
\par \vskip 0.25cm \indent
    At the start of each beamtime, the background scattering pattern of the empty beamline instrument is collected. This is referred to as an air scatter, and can be used to spot any systematic offsets in the experimental data. The second and third collected scattering patterns are of the empty cell with the beam pointed through the middle of the glass capillary, and clipping the edge of one the current collectors respectively. Both these measurements are to determine the background for subtraction during analysis. The third measurement is taken in the case that the microfocus beam is slightly misaligned and intersects one of the current collectors by accident.
\par \vskip 0.25cm \indent
    A custom heater has been designed for the DRIX cell, allowing studies to be performed at temperatures up to 180\(^\circ\)C for one cell at a time. This will open the possibility to further solid-state battery research.
\par
\subsubsection*{Applications of the DRIX cell}
    As mentioned previously, XPDF is mostly utilised when probing materials locally. It is therefore especially suitable for materials that lack a long-range order. For example, DRIX cells have been used to study disordered rock salts to understand whether they can successfully increase the charge capacity of lithium ion batteries \cite{diaz-lopez_rock_salts}.
\par \vskip 0.25cm \indent
    Although it is designed for the XPDF technique, it can also be used for other forms of X-ray spectroscopy, including techniques such as XAS and XRS. By using a DRIX cell, the signal may be optimised during XRS experiments. Its radial geometry allows for improvements in signal-to-noise ratio, as well as the ability to alter the sample thickness by adjusting the position of the beam on the capillary. Diaz-Lopez et al. comment that such experiments using the DRIX cell could lead to advances in bulk material characterisation \textit{in operando} \cite{diaz-lopez}.
\par \vskip 0.25cm \indent
    The small size limit of the sample and radial geometry mean that the TRL of the cell is very low, and results are not likely to be representative of a commercial cell. As such, the DRIX cell is most appropriate for fundamental materials research.
\subsubsection*{Known limitations of the DRIX cell}
    Synthesis of the DRIX cell is typically carried out on-site by the visiting users in one of the gloveboxes available at Diamond. However, the delicate nature of the small glass capillaries makes them fragile and difficult to assemble in the glovebox. Users are encouraged to arrive at least a day prior to their beamtime to assemble their cells, and to book out the glovebox in advance as their availability is limited. Furthermore, the low yield stress of the thin capillaries makes the cell unsuited to experiments under pressure.
\par \vskip 0.25cm \indent
    Radiation damage is usually insignificant during experiments on I15-1, due to the high energy of the incident X-rays. However, the beam may be modulated using its auto-filter regime. The flux can be adjusted from 0.001\% to 100\% in logarithmic increments.
\par

\newpage
\subsection*{Spectro-electrochemical flow cell [Santosh Kumar]}
\subsubsection*{Overview of the spectro-electrochemical flow cell}
    The spectro-electrochemical flow cell was created to allow rapid sample change when conducting soft XPS experiments. Its modular design allows the cell to be highly adaptable to different experimental set-ups. Whilst it has predominantly been used at the B07 beamline of Diamond, it has additionally been implemented on B22, B18, and the P22 beamline at the DESY synchrotron.
\par
\subsubsection*{Design of the spectro-electrochemical flow cell}
    Readers are referred to Kumar et al.'s paper for a more detailed analysis of the spectro-electrochemical flow cell design \cite{kumar}.
\par \vskip 0.25cm \indent
    The main cell body is constructed from PEEK, chosen for its material stability under a wide range of pH levels, as well as its ability to withstand vacuum conditions. It is 45mm in diameter, which is also the maximum size for the sample. Three electrodes can be connected into the cell body, although for battery research experiments the use of only two electrodes is more typical, and the third is covered with a blank. Liquid or gas may be flowed through to the cell via ID PEEK or PTFE tubing, thus allowing steady circulation of electrolyte solution.
\par
\begin{figure}[h]
    \centering
    \includegraphics[scale=0.25]{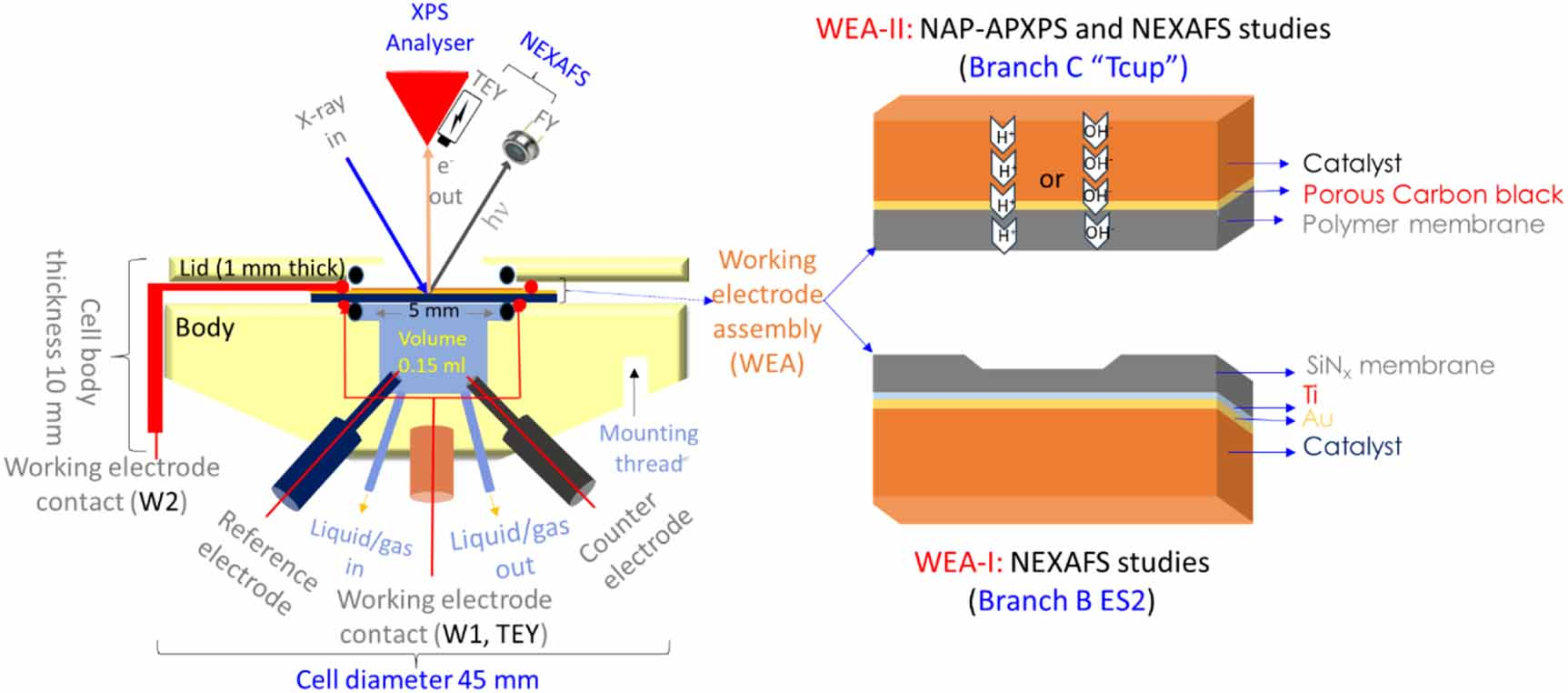}
    \caption{Diagram of the spectro-electrochemical flow cell, taken from Figure 2 of Kumar et al.'s paper \cite{kumar}}
    \label{fig:spectro_flow_fig}
\end{figure}

 \vskip 0.25cm \indent
    The working electrode assembly of the flow cell contains a thin, replaceable membrane which can be constructed from a variety of materials. Silicon nitride is often implemented as a window due to its transparency to X-rays, and is best suited for XAS applications. It must be under 200nm in thickness to allow for the transmission of the low-energy incident X-rays, and is usually of around 100nm in thickness. It is also possible to use mono-layer graphene windows; indeed, they are particularly suitable when carrying out XPS measurements. Electrons are able to pass through this single graphene layer, and can be subsequently be detected. Water-permeable polymer membranes are also a common choice - this includes materials such as Nafion\texttrademark{}. These different configurations are shown in Figure~\ref{fig:spectro_flow_fig}, where WEA-I depicts a working electrode assembly using a silicon nitride window, and WEA-II displays an assembly which uses a permeable polymer membrane.
\par
\subsubsection*{Experimental process using the spectro-electrochemical flow cell}
    As B07 is a soft X-ray beamline, the beam energy is relatively low. To ensure that there is minimal attenuation of the X-ray beams prior to reaching the sample, the chamber is evacuated during experiments.
\par \vskip 0.25cm \indent
    The cell can be cycled in both a static and flow state. However, the flow state is often preferred under X-ray irradiation since the constant exchange of electrolyte has a multifold benefit: the inner temperature and pH conditions remain constant, and any liquid or gas products may be released instead of accumulating inside the cell body cavity. Furthermore, the effects of beam damage can be alleviated, leading to higher reliability of results.
\par \vskip 0.25cm \indent
    One of the most important considerations is the flow rate of the chosen fluid through the cell. Although a high flow rate would minimise effects caused by radiation damage, a low flow rate is crucial to minimise flow-induced vibrations. This will in turn reduce artefacts in the experimental data, and ultimately result in a higher signal-to-noise ratio. A typical flow rate would be 2-4ml per minute, but a range of 2\(\mu\)l-20ml per minute can be achieved using the spectro-electrochemical flow cell. This consistent cycling of fluid also helps maintain the stack pressure in the cell.
\par
\subsubsection*{Applications of the spectro-electrochemical flow cell}
    The two branches of the versatile soft X-ray (VerSoX) beamline, B07, may operate simultaneously and independently. Branch B allows for XPS measurements to be taken under ultra-high vacuum (UHV), but also accommodates XANES experiments at ambient pressure or UHV. On B07, XAS is measured in two different detection modes: total electron yield (TEY), and total fluorescent yield (TFY). Whilst the former is more surface sensitive, the latter is suited for bulk measurements. On the other hand, branch C specialises in ambient pressure XPS \cite{grinter}. These ambient conditions are crucial for electrocatalysis, but are rarely achievable under conventional laboratory XPS set-ups. Hence, the flow cell opens up research possibilities regarding surface and interface analysis using an \textit{in operando} cell.
\par \vskip 0.25cm \indent
    The performance of the flow cell has been validated both off-beamline and \textit{in operando} through evaluating a series of electrochemical measurements and TEY/TFY signals \cite{kumar}. The results indicate that it can be used as an accurate and reliable tool for fundamental materials research at a low TRL, but it is not comparable to conventional commercial cells.
\subsubsection*{Known limitations of the spectro-electrochemical flow cell}
    For soft X-ray experiments in transmission mode, the sample must be adequately thin to allow for a sufficient proportion of the incident beam to penetrate through the cell. This can be a challenge to achieve, and may also hamper the quality of results due to a poor signal-to-noise ratio.
\par \vskip 0.25cm \indent
    For ambient XPS, sometimes the cap of the flow cell is removed to allow more electrons to be reflected back out and subsequently detected. However, this can cause the electrolyte flowing through the cell to dry out. To alleviate this issue, graphene windows are used. Some users may be hindered by the difficulty and cost of graphene preparation, which may limit the scope of experiments they can perform.
\par \vskip 0.25cm \indent
    Beam damage must still be taken into consideration as soft X-rays may interact significantly with the sample. Hence, it is necessary to first carry out control measurements during the beamtime to establish different artefact defects. This can also help identify issues with electrical connection, or flow rate. The beam can then be defocused to limit the radiation damage experienced.
\par \vskip 0.25cm \indent
    Different samples can be exchanged rapidly during the operation of the cell. The flow cell is compatible with both aqueous and organic electrolyte solutions of a wide range of pH. Theoretically, different temperatures of electrolyte may be cycled through the cell but there have not been any experiments to investigate such effects at of the time of writing this report.
\par

\newpage
\subsection*{\textit{In operando} electrochemical cells for B18 [Veronica Celorrio]}
\subsubsection*{Overview of the B18 \textit{in operando} cells}
    The B18 beamline supports several operando electrochemical cell configurations that provide progressively greater control over mass transport and gas management. The simplest arrangement is a static cell intended for stable measurements with minimal experimental complexity \cite{genovese}. Building on this, a flow cell introduces controlled electrolyte circulation to enable solution exchange and to mitigate bubble related artefacts \cite{wise}. The most specialised design is the SPEC-XAS spectro-electrochemical cell, developed to support gas-evolving and gas-consuming electrocatalysis studies by dedicated gas and electrolyte compartments \cite{sherwin}. Detailed engineering drawings and assembly information are provided in the cited publications and their supplementary material. 
    \par
\subsubsection*{Design of the B18 \textit{in operando} cells}
\begin{figure}[h]
    \centering
    \includegraphics[scale = 0.25]{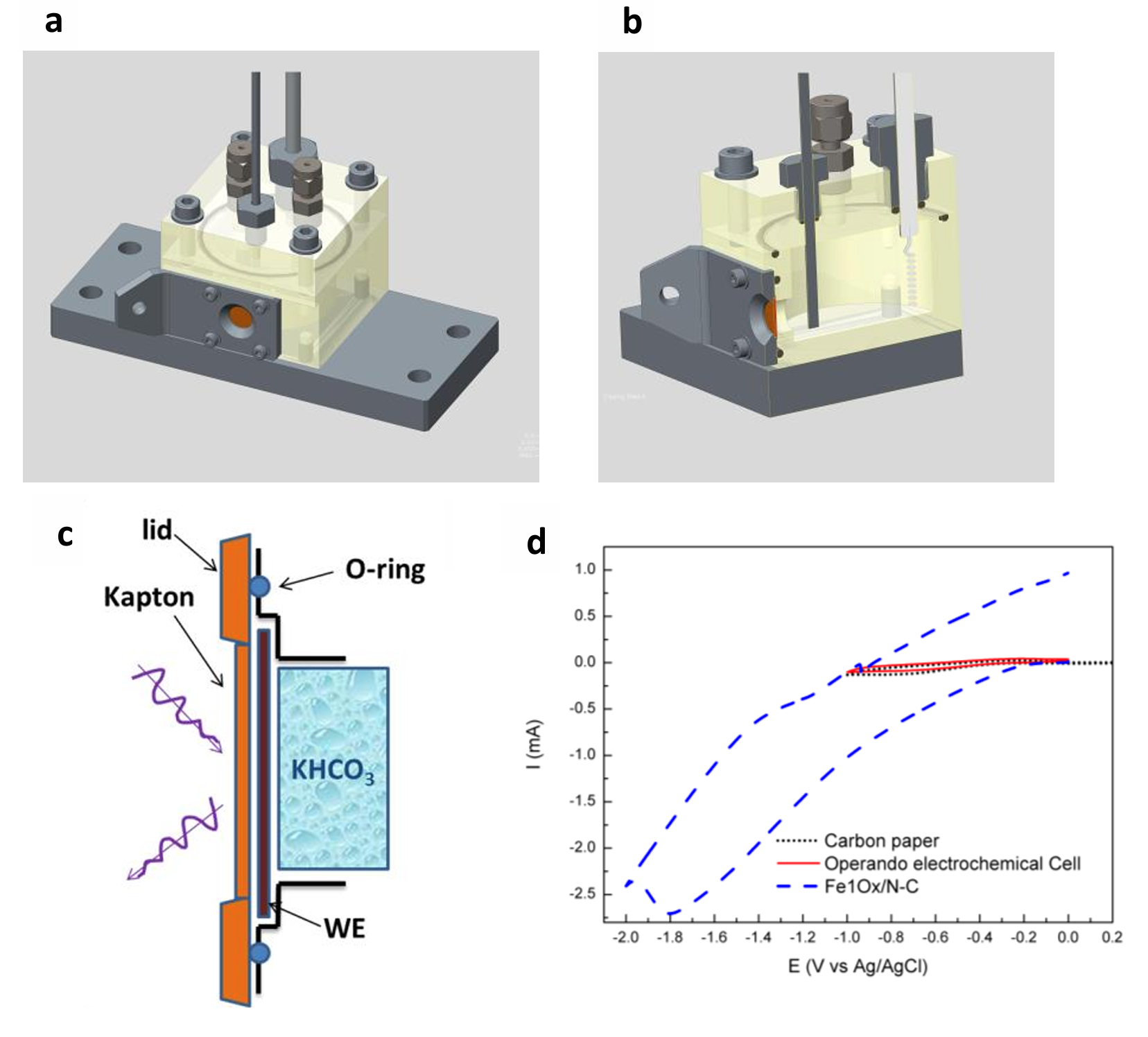}
    \caption{Schematics and graphs of the B18 static electrochemical cell, taken from Supplementary Figure 2 of Genovese et al.'s paper \cite{genovese}}
    \label{fig:B18_static}
\end{figure}

    The static electrochemical cell compromises a PTFE body incorporating three electrodes: the working electrode (WE), counter electrode (CE), and the reference electrode (RE). As seen from the labelled schematic in Figure~\ref{fig:B18_static}c, the WE sample is positioned so that it is in contact with the electrolyte while being separated from the incident X-ray beam by a Kapton window in the cell lid. In this case, the sample is facing the inside of the cell to keep contact with the electrolyte. The lid of the cell also functions as the WE current collector and is therefore commonly fabricated of a conductive material (vitreous carbon or stainless steel). To improve measurement quality, the cell geometry is designed to minimise the thickness of electrolyte between the WE and the Kapton window; careful selection of the recess depth reduces absorption and scattering from the electrolyte and improves the signal-to-noise ratio.
\par
\begin{figure}[h]
    \centering
    \includegraphics[scale = 0.2]{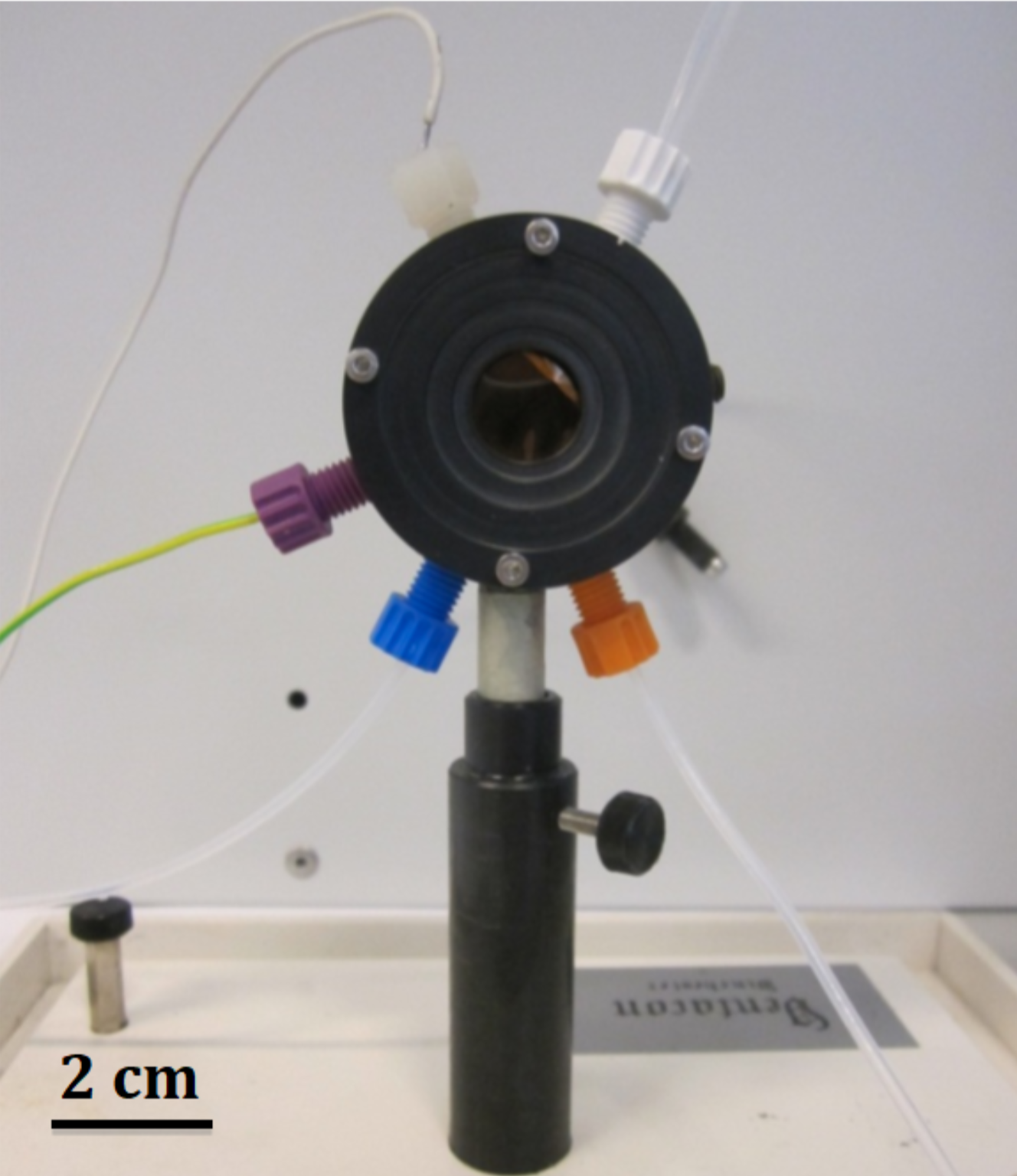}
    \caption{Photograph of the B18 static electrochemical cell, taken from Supplementary Figure 1 of Wise et al.'s paper \cite{wise}}
    \label{fig:B18_flow}
\end{figure}

\indent
    The flow cell extends the static design by enabling electrolyte circulation. Its body is manufactured from 30\(\%\) glass-filled polyphenyl sulphide (PPS), providing chemical resistance across a broad pH range and compatibility with both organic and aqueous electrolytes. Electrical connection of the WE is typically made through a gold current collector, whereas there is an option for platinum or titanium wires as the CE. Two fluidic ports are provided: an inlet for electrolyte inflow, and an outlet for electrolyte outflow. The outlet also functions as vent for gases generated at the WE, which is particularly important for gas-evolving electrocatalysis. As in the static cell, O-rings are used to maintain a hermetic seal and the X-ray windows are typically Kapton. A photo of the assembled flow cell is shown in Figure~\ref{fig:B18_flow}.
\par
\begin{figure}[h]
    \centering
    \includegraphics[scale = 0.2]{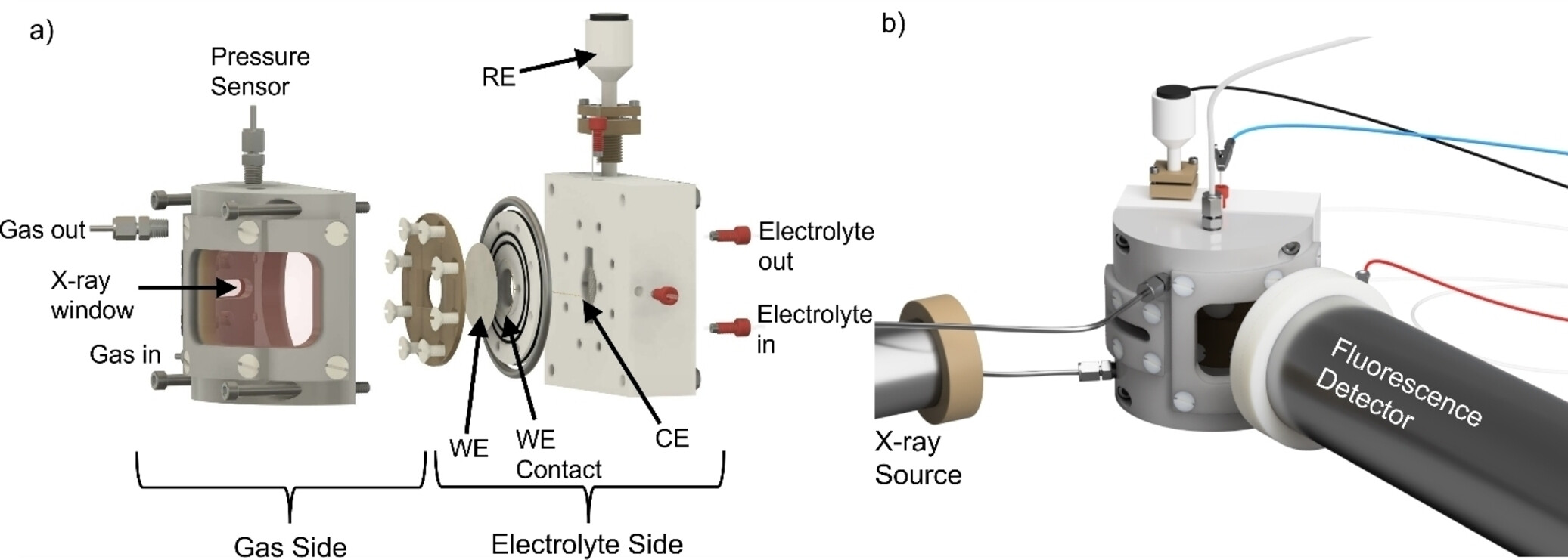}
    \caption{Schematics of the B18 SPEC-XAS cell, taken from Figure 1 of Sherwin et al.'s paper \cite{sherwin}}
    \label{fig:B18_specxas}
\end{figure}

\indent
    The SPEC-XAS cell introduces a two-compartment architecture. The incident X-ray beam enters the hemicylindrical gas-side where the WE is placed in the back and enables gas exchange. This configuration permits a range of incidence angles, typically between 15\(^\circ\) and 60\(^\circ\), allowing experimental flexibility and, potentially, improved signal collection. The second, cuboidal electrolyte-side houses the electrolyte, WE, CE and RE. This compartment incorporates a flow channel designed to support electrolyte circulation to control mass transport and assist with gas bubbles generation. The effective management of mass transport and bubbles are essential to produce high quality data and maintain catalyst utilisation and activity/turn over at current densities comparable to the optimised cell designs used in practical applications.The overall design of the B18 SPEC-XAS cell is depicted in Figure~\ref{fig:B18_specxas}.
\par
\subsubsection*{Experimental process using the B18 \textit{in operando} cells}
    Prior to the beamtime, users are encouraged to simulate the total absorption of all the relevant layers in their cells to ensure that the signal-to-noise ratio is adequate. If the data is being collected in fluorescence mode, the absorption of the pre-sample layers should be measured, whereas if the data is being collected in transmission mode, the total absorption of the cell must be found. The cells at B18 are optimised for detection in fluorescence mode, except the flow cell which also allows for transmission collection. To find the absorption of samples, Klemnetiev and Chernikov's XAFSmass program can be used. Subsequently, the Filter Transmission web page created by Lawrence Berkeley National Laboratory may be used to find the attenuation of the other materials present in the cell.
\par \vskip 0.25cm \indent 
    Hard X-ray exposure can induce radiolysis or other forms of radiation damage that alter the sample or electrolyte and compromise data interpretation. A common approach is to collect preliminary XAS spectra at open circuit potential for an extended period (approximately 30 minutes), both with the sample dry and with electrolyte present. Agreement between spectra supports the conclusion that beam-induced changes are negligible under the chosen conditions. If changes are observed, beam attenuation using filters can be employed to reduce dose and stabilise the measurement. 
\par \vskip 0.25cm \indent
    For the flow and SPEC-XAS cells, electrolyte circulation or solution switching are typically achieved using either a peristaltic or syringe pump. Continuous flow can reduce artefacts caused by bubble accumulation at the WE during gas evolution, although static operation remains possible when flow is not required. 
\par \vskip 0.25cm \indent
    As seen from Figure~\ref{fig:B18_specxas}a, a differential pressure sensor can be used to maintain a stable pressure environment in the gas compartment of the SPEC-XAS cell during operation. In addition,  this sensor may provide diagnostic information, as unexpected pressure behaviour can indicate problems such as blockages leakage of electrolyte into the gas diffusion channels \cite{sherwin}.
\par
\subsubsection*{Applications of the B18 \textit{in operando} cells}
    The B18 operando cells are used to quantify how electrode materials evolve under applied potential in environments that progressively increase in experimental control and relevance to gas-involving electrocatalysis. The static cell provides a low-complexity platform for establishing baseline behaviour and for conducting stable holds or stepwise protocols where a fixed electrolyte volume and a compact geometry are advantageous. The flow cell builds on this by enabling controlled electrolyte circulation, which supports electrolyte exchange experiments, improves reproducibility through better mass-transport control, and mitigates bubble accumulation during gas-evolving reactions. The SPEC-XAS cell is intended for the most demanding gas-evolving or gas-consuming studies. Across all three designs, the cells are best viewed as bespoke research platforms for mechanistic studies rather than as representations of commercial electrochemical devices, and the resulting observations should be interpreted within that scope. 
\par
\subsubsection*{Limitations of the B18 \textit{in operando} cells}
    A key practical constraint across the B18 in operando cell portfolio is the allowable detection geometry. Of the three designs described here, only the flow cell supports both transmission and fluorescence XAS measurements. In contrast, the static cell and the SPEC-XAS cell are configured for fluorescence measurements only, which restricts experimental options where transmission geometry would otherwise be preferable (for example, for highly concentrated samples or for minimising self-absorption effects).
Users must also account for radiation-induced damage when operating under hard X-ray illumination. Sustained exposure can alter the sample and/or degrade polymeric components and windows. Accordingly, beam effects should be evaluated through stability measurements at open circuit potential, and the incident intensity should be reduced using attenuation filters where necessary. Window material and thickness should be selected to balance mechanical robustness under prolonged irradiation against X-ray transmission and overall signal quality.
\par

\newpage
\subsection*{Modular \textit{Operando} Cells for Battery Research at I14 [Connor Wright; Miguel Gomez-Gonzalez]}

\subsubsection*{Overview of the M3.0 and M4.0 Cells}
The M3.0 and M4.0 cells are the third and fourth iterations of a modular \textit{operando} cell for battery research designed initially for use at Diamond Light Source’s I14 Nanoprobe beamline. They enable users to perform nanoscale spectroscopy, diffraction, and imaging of air-sensitive electrochemical materials. The aim is to improve the accuracy of correlations between multi-modal datasets by keeping the electrochemical environment constant across all modular setups.

\subsubsection*{Design of the M3.0 and M4.0 Cell}
\begin{figure}[h]
    \centering
    \includegraphics[scale = 1]{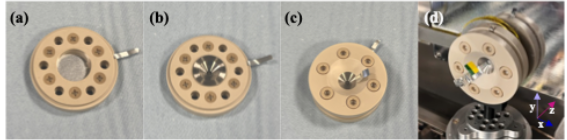}
    \caption{Showing (a-c) the assembly process of the M3.0 cell, and (d) the cell mounted onto I14 with the electrode stack visible through the optical window.}
    \label{fig:m3_assembly}
\end{figure}

Both the M3.0 and M4.0 cells are designed to be easy for non-experts to make. Like many other operando battery cells, they accept coin-type electrode stacks. The M3.0 cell consists of four main parts. The front and housing pieces screw together to secure the window and the electric contact in place (Figure~\ref{fig:m3_assembly}a). This can be completed outside of a glovebox. The remaining parts should be moved into the glovebox, and the electrode stack inserted as would be done for a coin cell before the module is inserted (Figure~\ref{fig:m3_assembly}b), and the back secured (Figure~\ref{fig:m3_assembly}c). All parts, excluding the stainless steel ‘modules’, are made from PEEK due to its chemical resilience and mechanical rigidity. A quirk of I14 is that very precise z-direction focusing is required. As a result, the ‘blank’ module – used for XANES and XRF mapping in fluorescence mode – requires an optically transparent window to allow optical focusing from the ‘back’, directly onto the electrode stack (Figure~\ref{fig:m3_assembly}d). This is achieved by a 5mm diameter, 1mm thick disk of Gorilla\textregistered Glass, secured in place via epoxy (Torr Seal\textregistered, Aligent). Electrical contacts are formed via pressure contacts onto pouch-cell-type tabs. These are mechanically rolled to reduce their thickness. No sealing issues have been found when the contacts are rolled adequately thin.

\begin{figure}[h]
    \centering
    \includegraphics[scale = 1]{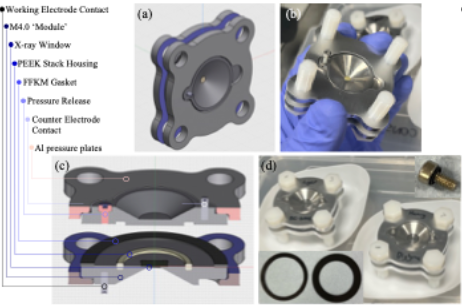}
    \caption{Showing (a and b) CAD and digital images of the M4.0 cell with an SiN window in place for tender X-ray analysis at B18. (c) cross-sectional CAD diagram, and (d) images of the ‘blank module’ backs, with (insert bottom) two different FFKM gaskets, and (insert top) the ‘pressure release’ O-ring sealing screw.}
    \label{fig:m4}
\end{figure}

The M4.0 cell (Figure~\ref{fig:m4}) is designed similarly to the M3.0, accepting coin-cell-type electrode stacks, but instead of a piston-type module providing sealing and pressure regulation, these functions are provided by an FFKM gasket that can range in thickness (0.8mm-2.0mm) and internal diameters (Figure~\ref{fig:m4}d, insert bottom). There is an in-built overhang of 0.4mm in the centre of all ‘back’ modules. Gaskets should be selected based on the electrode stack's pressure requirements and the coating/stack thickness. Smaller internal diameters will allow less pressure to be exerted onto the stack whilst maintaining adequate sealing. The pressure release addresses an issue found with the M3.0 cell, as explained in the following sections, whilst the large-diameter screws allow torque wrenches to be used for reproducible sealing. Electrical contacts are made via direct screws into the stainless steel modules.

\subsubsection*{Experimental Process using the M3.0 and M4.0 Cell}
The first item on any user’s agenda should be to determine which windows and cell ‘module’ would be needed for their chosen experiment and beamline. For instance, choosing to work with SiN windows should be done only if necessary, given their high cost and extreme brittleness. Thin (~0.1mm) PEEK disks can also be used if attenuation is less of a problem. The module designs will be available to request in advance from I14, allowing modifications to accommodate different window fittings, provided there is enough time for manufacture before the user’s beamtime. Holders for the M3.0 and M4.0 have been developed to be compatible with the I14 and B18 endstations, with other mounts to be available in the future.

It is necessary to investigate the effects of beam damage on any system under investigation. The long run times and the concentrated, high-flux required to obtain nano-spatially resolved XANES data, for example, will result in significant radiation doses to the sample, even at high energies. Repeated measurements can usually achieve this whilst the battery is under open-circuit conditions (i.e. no electrochemistry running). If no changes are measured over the same time period as the active investigation, beam damage can be ignored.

\subsubsection*{Applications of the M3.0 Cell}
These cells have been designed specifically to comply with the I14 endstation. Other beamlines are less stringent, except for reduced-pressure endstations (e.g., B07). The primary use case for these cells is spectromicroscopy of battery electrode materials in fluorescence modes, with the M3.0 and M4.0 being used for such experiments on I14 and B18 beamlines, respectively, in the past. The hope is to extend these use cases to include transmission modalities, such as with the TXM-XRD capabilities of I14. The switch from fluorescence to transmission modes is simple for these cells, with only the modules needing to be replaced (in the case of the M4.0), or the module and the front PEEK piece (for the M3.0). Assembly remains the same, although modifications to the electrode stack may be necessary, such as the use of a ring-shaped counter electrode \cite{frank2024aurex}.

Unlike some electrochemical operando cells, the M4.0 cell, in particular, is designed with representative electrochemistry at the forefront, emphasising reproducible and high stack pressures, a limitation of most liquid-based operando cells \cite{borkiewicz2015best}.

\subsubsection*{Known Limitations of the M3.0 Cell}

\begin{figure}[h]
    \centering
    \includegraphics[scale = 1]{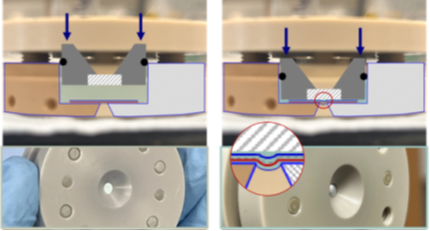}
    \caption{Schematic overlays onto the M3.0 cell assembly, highlighting the bulging effect when fully tightened.}
    \label{fig:m3_bulging}
\end{figure}

The plunger-type ‘module’ and its O-ring create a very tight seal from the atmosphere, but can also lead to bulging of the flexible Al-pouch window from trapped excess gas on assembly (Figure~\ref{fig:m3_bulging}). This leads to a trade-off between electrode stack uniformity and overall stack pressure. Negating bulging is important for representative electrochemistry of the probed area \cite{borkiewicz2015best}.

For this reason, the latest design – so-called the M4.0 cell – moves away from the piston-type O-ring and module in favour of an FFKM gasket. This allows the inclusion of a ‘pressure release’ hole that can be left open during sealing/pressurising and subsequently sealed, reducing the risk of bulging and enabling the use of SiN windows with a much reduced failure rate. However, the M4.0 cell is not without issue. Currently, the use of aluminium pressure plates can leave the cell prone to accidental electrical shorting. In the future, this will be fixed by non-conductive interlayers or painting. A further limitation is the FFKM gaskets. There is relatively poor availability of such gaskets, limiting the thicknesses available to users. For more choice and, ultimately, better pressure and sealing, a downgrade to less chemically resistant FKM (Viton) O-rings is often necessary.
        
\newpage
\subsection*{Coin and pouch cells [Scott Young, Gabriel Perez, Sylvia Britto, Isabel Antony]}\label{sec:CoinPouch}
\subsubsection*{Overview}
While many bespoke cell designs are provided by different beamlines and instruments across facilities, users frequently bring laboratory-assembled cells, such as standard coin or pouch cell formats, to maintain continuity with their home laboratory data. However, adapting these standard geometries for \textit{operando} experiments requires specific modifications to balance electrochemical integrity with the requirements of beam transmission and scattering geometry.

\subsubsection*{Coin Cells} 
Coin cells remain a standard form factor across electrochemical laboratories. However, their stainless steel casings are highly attenuating for X-rays and neutrons. To enable beam penetration, the casings must often be thinned or replaced with alternative window materials (e.g., Kapton or thinner metals). These modifications can introduce non-uniform pressure distributions, which ultimately affect electrical contact resistance and electrochemical performance compared to an unmodified standard cell \cite{borkiewicz2015best}.

\subsubsection*{Pouch Cells} Pouch cells offer a viable alternative, as their thinner polymer-aluminium casings typically permit higher X-ray and neutron transmission. However, hand-assembled pouch cells generally exhibit lower reproducibility than coin cells under standard testing conditions.

For neutron techniques, specifically on instruments like NIMROD and POLARIS, single-layer pouch cells have been successfully measured. The experimental protocol typically involves acquiring a background measurement of an empty pouch cell, which is subsequently subtracted from the operando data to isolate the signal from the active materials. Pouch cells have also been tested successfully on the EMU muon spectrometer.

\subsubsection*{Limitations and Holders} 
A critical challenge when using pouch cells on instruments like POLARIS or in vacuum environments is maintaining adequate stack pressure. Vacuum conditions can cause pouch casings to expand or deform, potentially reducing electrical contact and compromising the electrochemical circuit \cite{borkiewicz2015best}.

To mitigate this, custom sample environments are often required. For example, Parks et al. \cite{parks2025non} implemented a custom-made pouch cell holder fabricated from Polyether Ether Ketone (PEEK). This holder incorporated an aluminium window to mechanically compress the pouch cell during the experiment, ensuring consistent stack pressure while simultaneously maximising X-ray transmission. Similar clamping mechanisms are recommended for neutron experiments to ensure reproducibility and prevent delamination of electrode stacks under vacuum.

%% file: sample.bib
@manual{smith,
    author = "Smith, R. I. and Hull, S",
    title = "User Guide for Polaris Powder Diffractometer at ISIS",
    url = "https://www.isis.stfc.ac.uk/Pages/user-guide-for-the-polaris-powder-diffractometer-at-isis.pdf",
    addendum = "(accessed: 23.07.2025)"
}

@article{biendicho,
    title = {New in-situ neutron diffraction cell for electrode materials},
    journal = {Journal of Power Sources},
    volume = {248},
    pages = {900--904},
    year = {2014},
    issn = {0378-7753},
    doi = {https://doi.org/10.1016/j.jpowsour.2013.09.141},
    url = {https://www.sciencedirect.com/science/article/pii/S0378775313016492},
    author = {Biendicho, Jordi Jacas and Roberts, Matthew and Offer, Colin and Noréus, Dag and Widenkvist, Erika and Smith, Ronald I. and Svensson, Gunnar and Edström, Kristina and Norberg, Stefan T. and Eriksson, Sten G. and Hull, Stephen},
}

@article{taminato,
    author = {Taminato, Sou and Yonemura, Masao and Shiotani, Shinya and Kamiyama, Takashi and Torii, Shuki and Nagao, Miki and Ishikawa, Yoshihisa and Mori, Kazuhiro and Fukunaga, Toshiharu and Onodera, Yohei and Naka, Takahiro and Morishima, Makoto and Ukyo, Yoshio and Adipranoto, Dyah Sulistyanintyas and Arai, Hajime and Uchimoto, Yoshiharu and Ogumi, Zempachi and Suzuki, Kota and Hirayama, Masaaki and Kanno, Ryoji},
    title = {Real-time observations of lithium battery reactions—operando neutron diffraction analysis during practical operation},
    journal = {Scientific Reports},
    year = {2016},
    doi = {https://doi.org/10.1038/srep28843},
    volume = {6}
}

@article{srinivasan,
    author = {Srinivasan, R. K. and Chandran, S. Ravi and Chen, Y. and An, K.},
    title = {In-Operando Neutron Diffraction Investigation of Structural Transitions during Lithiation of Si Electrode in Li-Ion Battery},
    journal = {The Electrochemical Society},
    year = {2022},
    doi = {10.1149/1945-7111/ac9a7e},
}

@article{perez,
    author = {Pérez, Gabriel and Brittain, Jake and McClelland, Innes and Hull, Stephen and Jones, Martin and Playford, Helen and Cussen, Serena and Baker, Peter and Reynolds, Emily},
    year = {2023},
    month = {01},
    title = {Neutron and Muon Techniques for Battery Materials Studies},
    volume = {11},
    journal = {Journal of Materials Chemistry A},
    doi = {10.1039/D2TA07235A}
}

@phdthesis{mcclelland,
    author = "McClelland, Innes and Cussen, Serena A. and Baker, Peter J.",
    title = "Towards Operando Measurements of Li+ Diffusion in Batteries using Muon Spectroscopy",
    school = "The University of Sheffield",
    year = "2022",
    month = "03"
}

@article{reynolds,
    author = {Reynolds, Emily and Fitzpatrick, Jack and Jones, Martin and Tapia-Ruiz, Nuria and Playford, Helen and Hull, Stephen and McClelland, Innes and Baker, Peter and Cussen, Serena and Pérez, Gabriel},
    journal = {Journal of Materials Chemistry A},
    year = {2024},
    month = {01},
    title = {Investigation of sodium insertion in hard carbon with operando small angle neutron scattering},
    volume = {12},
    doi = {10.1039/D3TA04739C},
}

@article{diaz-lopez,
    author = "Diaz-Lopez, Maria and Cutts, Geoffrey L. and Allan, Phoebe K. and Keeble, Dean S. and Ross, Allan and Pralong, Valerie and Spiekermann, Georg and Chater, Philip A.",
    title = "{Fast {\it operando} X-ray pair distribution function using the DRIX electrochemical cell}",
    journal = "Journal of Synchrotron Radiation",
    year = "2020",
    volume = "27",
    number = "5",
    pages = "1190--1199",
    month = "09",
    doi = {10.1107/S160057752000747X},
    url = {https://doi.org/10.1107/S160057752000747X},
}

@article{diaz-lopez_rock_salts,
author = {Diaz-Lopez, Maria and Chater, Philip A. and Bordet, Pierre and Freire, Melanie and Jordy, Christian and Lebedev, Oleg I. and Pralong, Valerie},
title = {{Li$_2$O:Li-Mn-O} Disordered Rock-Salt Nanocomposites as Cathode Prelithiation Additives for High-Energy Density Li-Ion Batteries},
journal = {Advanced Energy Materials},
volume = {10},
number = {7},
pages = {1902788},
doi = {https://doi.org/10.1002/aenm.201902788},
url = {https://advanced.onlinelibrary.wiley.com/doi/abs/10.1002/aenm.201902788},
year = {2020}
}

@article{drnec,
    author = {Drnec, Jakub and Lyonnard, Sandrine},
    journal = {Nature Nanotechnology},
    title = {Battery research needs more reliable, representative and reproducible synchrotron characterizations},
    url = {https://doi.org/10.1038/s41565-025-01921-4},
    year = {2025},
    month = {05},
    pages = {584--587},
    volume = {20},
    issue = {5}
}

@article{lin,
    author = {Lin, Feng and Liu, Yijin and Yu, Xiqian and Cheng, Lei and Singer, Andrej and Shpyrko, Oleg G. and Xin, Huolin L. and Tamura, Nobumichi and Tian, Chixia and Weng, Tsu-Chien and Yang, Xiao-Qing and Meng, Ying Shirley and Nordlund, Dennis and Yang, Wanli and Doeff, Marca M.},
    title = {Synchrotron X-ray Analytical Techniques for Studying Materials Electrochemistry in Rechargeable Batteries},
    journal = {Chemical Reviews},
    volume = {117},
    number = {21},
    pages = {13123-13186},
    year = {2017},
    doi = {10.1021/acs.chemrev.7b00007},
    note ={PMID: 28960962},
    URL = {https://doi.org/10.1021/acs.chemrev.7b00007},
}

@article{bras,
    author = {Bras, Wim and Myles, Dean and Felici, Roberto},
    year = {2021},
    month = {07},
    title = {When X-rays alter the course of your experiments},
    volume = {33},
    journal = {Journal of Physics: Condensed Matter},
    doi = {10.1088/1361-648X/ac1767}
}

@article{mcclelland_diffusion,
    author = {McClelland, Innes and Booth, Samuel G. and Anthonisamy, Nirmalesh N. and Middlemiss, Laurence A. and Perez, Gabriel E. and Cussen, Edmund J. and Baker, Peter J. and Cussen, Serena A.},
    title = {Direct Observation of Dynamic Lithium Diffusion Behavior in Nickel-Rich, LiNi0.8Mn0.1Co0.1O2 (NMC811) Cathodes Using Operando Muon Spectroscopy},
    journal = {Chemistry of Materials},
    volume = {35},
    number = {11},
    pages = {4149-4158},
    year = {2023},
    doi = {10.1021/acs.chemmater.2c03834},
    URL = {https://doi.org/10.1021/acs.chemmater.2c03834},
}

@article{kumar,
    doi = {10.1088/2515-7655/ad54ee},
    url = {https://dx.doi.org/10.1088/2515-7655/ad54ee},
    year = {2024},
    month = {06},
    publisher = {IOP Publishing},
    volume = {6},
    number = {3},
    pages = {036001},
    author = {Kumar, Santosh and Counter, James J C and Grinter, David C and Spronsen, Matthijs A Van and Ferrer, Pilar and Large, Alex and Orzech, Marcin W and Jerzy Wojcik, Pawel and Held, Georg},
    title = {An electrochemical flow cell for operando XPS and NEXAFS investigation of solid–liquid interfaces},
    journal = {Journal of Physics: Energy}
    }

@incollection{blundell,
    author = {Blundell, Stephen J. and De Renzi, Roberto and Lancaster, Tom and Pratt, Francis L.},
    isbn = {9780198858959},
    title = {The basics of µSR},
    booktitle = {Muon Spectroscopy: An Introduction},
    publisher = {Oxford University Press},
    year = {2021},
    month = {11},
    doi = {10.1093/oso/9780198858959.003.0001},
    url = {https://doi.org/10.1093/oso/9780198858959.003.0001},
    eprint = {https://academic.oup.com/book/0/chapter/366295851/chapter-pdf/50190692/oso-9780198858959-chapter-1.pdf},
}

@article{matsubara,
    author = {Matsubara, Nami and Nocerino, Elisabetta and Forslund, Ola Kenji and Zubayer, Anton and Papadopoulos, Konstantinos and Andreica, Daniel and Sugiyama, Jun and Palm, Rasmus and Guguchia, Zurab and Cottrell, Stephen P. and Kamiyama, Takashi and Saito, Takashi and Kalaboukhov, Alexei and Sassa, Yasmine and Masese, Titus and Månsson, Martin},
    title = {Magnetism and ion diffusion in honeycomb layered oxide \(K_2 Ni_2 TeO_6\)},
    journal = {Scientific Reports},
    year = {2020},
    month = {10},
    volume = {10},
    issue = {1},
    url = {https://doi.org/10.1038/s41598-020-75251-x},
    doi = {10.1038/s41598-020-75251-x}
}

@article{wright,
    doi = {10.1149/MA2025-013292mtgabs},
    url = {https://dx.doi.org/10.1149/MA2025-013292mtgabs},
    year = {2025},
    month = {07},
    publisher = {The Electrochemical Society, Inc.},
    volume = {MA2025-01},
    number = {3},
    pages = {292},
    author = {Wright, Zoe and Wood, Thomas and Perry McLean, Emma and Baboo, Joseph Paul and Reynolds, Emily and Baker, Peter J. and David, William I F},
    title = {Investigation of Na+ Diffusion in Prototypic Layered Transition Metal Oxide Cathode Materials Using Muon Spectroscopy},
    journal = {ECS Meeting Abstracts},
}

@article{ferdani,
    author ="Ferdani, Dominic W. and Pering, Samuel R. and Ghosh, Dibyajyoti and Kubiak, Peter and Walker, Alison B. and Lewis, Simon E. and Johnson, Andrew L. and Baker, Peter J. and Islam, M. Saiful and Cameron, Petra J.",
    title  ="Partial cation substitution reduces iodide ion transport in lead iodide perovskite solar cells",
    journal  ="Energy Environ. Sci.",
    year  ="2019",
    volume  ="12",
    issue  ="7",
    pages  ="2264-2272",
    publisher  ="The Royal Society of Chemistry",
    doi  ="10.1039/C9EE00476A",
    url  ="http://dx.doi.org/10.1039/C9EE00476A"}

@article{grinter,
    author = "Grinter, David C. and Ferrer, Pilar and Venturini, Federica and van Spronsen, Matthijs A. and Large, Alexander I. and Kumar, Santosh and Jaugstetter, Maximilian and Iordachescu, Alex and Watts, Andrew and Schroeder, Sven L. M. and Kroner, Anna and Grillo, Federico and Francis, Stephen M. and Webb, Paul B. and Hand, Matthew and Walters, Andrew and Hillman, Michael and Held, Georg",
    title = "{VerSoX B07-B: a high-throughput XPS and ambient pressure NEXAFS beamline at Diamond Light Source}",
    journal = "Journal of Synchrotron Radiation",
    year = "2024",
    volume = "31",
    number = "3",
    pages = "578--589",
    month = "05",
    doi = {10.1107/S1600577524001346},
    url = {https://doi.org/10.1107/S1600577524001346}
}

@article{thompson,
    author = {Thompson, S. P. and Parker, J. E. and Potter, J. and Hill, T. P. and Birt, A. and Cobb, T. M. and Yuan, F. and Tang, C. C.},
    title = {Beamline I11 at Diamond: A new instrument for high resolution powder diffraction},
    journal = {Review of Scientific Instruments},
    volume = {80},
    number = {7},
    pages = {075107},
    year = {2009},
    month = {07},
    issn = {0034-6748},
    doi = {10.1063/1.3167217},
    url = {https://doi.org/10.1063/1.3167217}
}

@article{sherwin,
    author = {Sherwin, Connor and Celorrio, Veronica and Podbevsek, Ursa and Rigg, Katie and Hodges, Toby and Ibraliu, Armando and Telfer, Abbey J. and McLeod, Lucy and Difilippo, Alessandro and Corbos, Elena C. and Zalitis, Chris and Russell, Andrea E.},
    title = {An optimised Cell for in situ XAS of Gas Diffusion Electrocatalyst Electrodes},
    journal = {ChemCatChem},
    volume = {16},
    number = {19},
    pages = {e202400221},
    doi = {https://doi.org/10.1002/cctc.202400221},
    url = {https://chemistry-europe.onlinelibrary.wiley.com/doi/abs/10.1002/cctc.202400221},
    year = {2024}
}

@article{wise,
    title = {Inhibitive effect of Pt on Pd-hydride formation of Pd@Pt core-shell electrocatalysts: An in situ EXAFS and XRD study},
    journal = {Electrochimica Acta},
    volume = {262},
    pages = {27-38},
    year = {2018},
    issn = {0013-4686},
    doi = {https://doi.org/10.1016/j.electacta.2017.12.161},
    url = {https://www.sciencedirect.com/science/article/pii/S0013468617327354},
    author = {Anna M. Wise and Peter W. Richardson and Stephen W.T. Price and Gaël Chouchelamane and Laura Calvillo and Patrick J. Hendra and Michael F. Toney and Andrea E. Russell}
}

@article{genovese,
    author = {Genovese, Chiara and Schuster, Manfred E. and Gibson, Emma K. and Gianolio, Diego and Posligua, Victor and Grau-Crespo, Ricardo and Cibin, Giannantonio and Wells, Peter P. and Garai, Debi and Solokha, Vladyslav and Krick Calderon, Sandra and Velasco-Velez, Juan J. and Ampelli, Claudio and Perathoner, Siglinda and Held, Georg and Centi, Gabriele and Arrigo, Rosa},
    title = {Operando spectroscopy study of the carbon dioxide electro-reduction by iron species on nitrogen-doped carbon},
    journal = {Nature Communications},
    year = {2018},
    volume = {9},
    issue = {1},
    url = {https://doi.org/10.1038/s41467-018-03138-7},
    doi = {10.1038/s41467-018-03138-7}
}

@article{saito,
    author = {Saito, Soshi and Watanabe, Hikari and Hayashi, Yutaka and Matsugami, Masaru and Tsuzuki, Seiji and Seki, Shiro and Canongia Lopes, Jos{\'e} N. and Atkin, Rob and Ueno, Kazuhide and Dokko, Kaoru and Watanabe, Masayoshi and Kameda, Yasuo and Umebayashi, Yasuhiro},
    title = {Li+ Local Structure in Li–Tetraglyme Solvate Ionic Liquid Revealed by Neutron Total Scattering Experiments with the 6/7Li Isotopic Substitution Technique},
    journal = {The Journal of Physical Chemistry Letters},
    volume = {7},
    number = {14},
    pages = {2832-2837},
    year = {2016},
    doi = {10.1021/acs.jpclett.6b01266},
    URL = {https://doi.org/10.1021/acs.jpclett.6b01266},
    eprint = {https://doi.org/10.1021/acs.jpclett.6b01266}
}

@article{ramadhan,
    author = "Ramadhan, Ranggi S. and Kockelmann, Winfried and Minniti, Triestino and Chen, Bo and Parfitt, David and Fitzpatrick, Michael E. and Tremsin, Anton S.",
    title = "{Characterization and application of Bragg-edge transmission imaging for strain measurement and crystallographic analysis on the IMAT beamline}",
    journal = "Journal of Applied Crystallography",
    year = "2019",
    volume = "52",
    number = "2",
    pages = "351--368",
    month = "04",
    doi = {10.1107/S1600576719001730},
    url = {https://doi.org/10.1107/S1600576719001730}
}

@article{headen,
    doi = {10.1149/MA2025-01462455mtgabs},
    url = {https://dx.doi.org/10.1149/MA2025-01462455mtgabs},
    year = {2025},
    month = {07},
    publisher = {The Electrochemical Society, Inc.},
    volume = {MA2025-01},
    number = {46},
    pages = {2455},
    author = {Headen, Thomas and Shutt, Rebecca and Shah, Ami and Gilani, Rakin and Clancy, Adam and Di Mino, Camilla and Cullen, Patrick and Skipper, Neal and Howard, Chris Anthony},
    title = {(Invited) Wide Q-Range Neutron Total Scattering: A Unique Tool for Understanding Electrochemical Interfaces},
    journal = {ECS Meeting Abstracts}
}

@article{busch,
    author = {Busch, Johanna and Niemann, Thomas and Neumann, Jan and Stange, Peter and Gärtner, Sabrina and Youngs, Tristan and Youngs, Sarah and Paschek, Dietmar and Ludwig, Ralf},
    title = {Role of Hydrogen Bond Defects for Cluster Formation and Distribution in Ionic Liquids by Means of Neutron Diffraction and Molecular Dynamics Simulations},
    journal = {ChemPhysChem},
    volume = {24},
    number = {12},
    pages = {e202300031},
    doi = {https://doi.org/10.1002/cphc.202300031},
    url = {https://chemistry-europe.onlinelibrary.wiley.com/doi/abs/10.1002/cphc.202300031},
    eprint = {https://chemistry-europe.onlinelibrary.wiley.com/doi/pdf/10.1002/cphc.202300031},
    year = {2023}
}

@article{christensen2023beam,
  title={Beam damage in operando X-ray diffraction studies of Li-ion batteries},
  author={Christensen, Christian Kolle and Karlsen, Martin Aaskov and Drejer, Andreas {\O}stergaard and Andersen, Bettina Pilgaard and Jakobsen, Christian Lund and Johansen, Morten and S{\o}rensen, Daniel Risskov and Kantor, Innokenty and J{\o}rgensen, Mads Ry Vogel and Ravnsb{\ae}k, Dorthe Bomholdt},
  journal={Synchrotron Radiation},
  volume={30},
  number={3},
  pages={561--570},
  year={2023},
  publisher={International Union of Crystallography}
}

@article{groves2025lithium,
  title={Lithium solvation and anion-dominated domain structure in water-in-salt electrolytes},
  author={Groves, Timothy S and Agg, Kieran J and Miao, Shurui and Headen, Thomas F and Youngs, Tristan GA and Smith, Gregory N and Perkin, Susan and Hallett, James E},
  journal={EES batteries},
  volume={1},
  number={6},
  pages={1797--1808},
  year={2025},
  publisher={Royal Society of Chemistry}
}

@article{parks2025non,
  title={Non-linear cracking response to voltage revealed by operando X-ray tomography in polycrystalline NMC811},
  author={Parks, Huw CW and Jones, Matthew P and Wade, Aaron and Llewellyn, Alice V and Tan, Chun and Reid, Hamish T and Ziesche, Ralf and Heenan, Thomas MM and Marathe, Shashidhara and Rau, Christoph and others},
  journal={EES Batteries},
  volume={1},
  number={3},
  pages={482--494},
  year={2025},
  publisher={Royal Society of Chemistry}
}

@article{frank2024aurex,
  title={The AUREX cell: a versatile operando electrochemical cell for studying catalytic materials using X-ray diffraction, total scattering and X-ray absorption spectroscopy under working conditions},
  author={Frank, Sara and Ceccato, Marcel and Jeppesen, Henrik S and Marks, Melissa J and Nielsen, Mads LN and Lu, Ronghui and Gammelgaard, Jens Jakob and Quinson, Jonathan and Sharma, Ruchi and Jensen, Julie S and others},
  journal={Applied Crystallography},
  volume={57},
  number={5},
  year={2024},
  publisher={International Union of Crystallography}
}

@misc{borkiewicz2015best,
  title={Best practices for operando battery experiments: influences of X-ray experiment design on observed electrochemical reactivity},
  author={Borkiewicz, Olaf J and Wiaderek, Kamila M and Chupas, Peter J and Chapman, Karena W},
  journal={The journal of physical chemistry letters},
  volume={6},
  number={11},
  pages={2081--2085},
  year={2015},
  publisher={ACS Publications}
}

@article{shah,
    doi = {10.1088/2515-7639/ac24ec},
    url = {https://dx.doi.org/10.1088/2515-7639/ac24ec},
    year = {2021},
    month = {sep},
    publisher = {IOP Publishing},
    volume = {4},
    number = {4},
    pages = {042008},
    author = {Shah, Ami R and Shutt, Rebecca R C and Smith, Keenan and Hack, Jennifer and Neville, Tobias P and Headen, Thomas F and Brett, Dan J L and Howard, Christopher A and Miller, Thomas S and Cullen, Patrick L},
    title = {Neutron studies of Na-ion battery materials},
    journal = {Journal of Physics: Materials}
}

@article{grey2017sustainability,
  title={Sustainability and in situ monitoring in battery development},
  author={Grey, CP and Tarascon, JM},
  journal={Nature materials},
  volume={16},
  number={1},
  pages={45--56},
  year={2017},
  publisher={Nature Publishing Group UK London}
}

@article{chapon2019diamond,
  title={Diamond-II: Conceptual Design Report},
  author={Chapon, LC and Boscaro-Clarke, I and Dent, AJ and Harrison, A and Launchbury, M and Stuart, DI and Walker, RP and others},
  journal={Diamond Light Source, Oxfordshire, UK},
  year={2019}
}

@article{messina2017exascale,
  title={The exascale computing project},
  author={Messina, Paul},
  journal={Computing in Science \& Engineering},
  volume={19},
  number={3},
  pages={63--67},
  year={2017},
  publisher={IEEE}
}

@article{stach2021autonomous,
  title={Autonomous experimentation systems for materials development: A community perspective},
  author={Stach, Eric and DeCost, Brian and Kusne, A Gilad and Hattrick-Simpers, Jason and Brown, Keith A and Reyes, Kristofer G and Schrier, Joshua and Billinge, Simon and Buonassisi, Tonio and Foster, Ian and others},
  journal={Matter},
  volume={4},
  number={9},
  pages={2702--2726},
  year={2021},
  publisher={Elsevier}
}

@article{burger2020mobile,
  title={A mobile robotic chemist},
  author={Burger, Benjamin and Maffettone, Phillip M and Gusev, Vladimir V and Aitchison, Catherine M and Bai, Yang and Wang, Xiaoyan and Li, Xiaobo and Alston, Ben M and Li, Buyi and Clowes, Rob and others},
  journal={Nature},
  volume={583},
  number={7815},
  pages={237--241},
  year={2020},
  publisher={Nature Publishing Group UK London}
}

@article{edge2021lithium,
  title={Lithium ion battery degradation: what you need to know},
  author={Edge, Jacqueline S and O’Kane, Simon and Prosser, Ryan and Kirkaldy, Niall D and Patel, Anisha N and Hales, Alastair and Ghosh, Abir and Ai, Weilong and Chen, Jingyi and Yang, Jiang and others},
  journal={Physical Chemistry Chemical Physics},
  volume={23},
  number={14},
  pages={8200--8221},
  year={2021},
  publisher={Royal Society of Chemistry}
}

@article{le2021x,
  title={X-ray tomography for lithium ion battery electrode characterisation—A review},
  author={Le Houx, James and Kramer, Denis},
  journal={Energy Reports},
  volume={7},
  pages={9--14},
  year={2021},
  publisher={Elsevier}
}

@article{le2025nanoscale,
  title={Nanoscale origins of chemomechanical degradation and failure in battery cathodes revealed by operando dark field X-ray microscopy},
  author={Le Houx, James and Mistry, Jessica and Spencer-Jolly, Dominic},
  journal={chemRxiv},
  year={2026},
}

@article{le2021openimpala,
  title={Openimpala: open source image based parallisable linear algebra solver},
  author={Le Houx, James and Kramer, Denis},
  journal={SoftwareX},
  volume={15},
  pages={100729},
  year={2021},
  publisher={Elsevier}
}

@article{menkin2024insights,
  title={Insights into soft short circuit-based degradation of lithium metal batteries},
  author={Menkin, Svetlana and Fritzke, Jana B and Larner, Rebecca and de Leeuw, Cas and Choi, Yoonseong and Gunnarsd{\'o}ttir, Anna B and Grey, Clare P},
  journal={Faraday Discussions},
  volume={248},
  pages={277--297},
  year={2024},
  publisher={Royal Society of Chemistry}
}

@article{ziesche2022neutron,
  title={Neutron imaging of lithium batteries},
  author={Ziesche, Ralf F and Kardjilov, Nikolay and Kockelmann, Winfried and Brett, Dan JL and Shearing, Paul R},
  journal={Joule},
  volume={6},
  number={1},
  pages={35--52},
  year={2022},
  publisher={Elsevier}
}

@inproceedings{bicer2017real,
  title={Real-time data analysis and autonomous steering of synchrotron light source experiments},
  author={Bicer, Tekin and Gursoy, Doga and Kettimuthu, Rajkumar and Foster, Ian T and Ren, Bin and De Andrede, Vincent and De Carlo, Francesco},
  booktitle={2017 IEEE 13th International Conference on e-Science (e-Science)},
  pages={59--68},
  year={2017},
  organization={IEEE}
}

@article{xu2025operando,
  title={Operando X-ray Computed Tomography Reveals the Role of Interfacial Nucleation Nanolayers in Suppressing Mechanical Failure in Zero-Excess Lithium All-Solid-State Batteries},
  author={Xu, Linfeng and Le Houx, James and Kachkanov, Vyacheslav and Zhang, Jinsong and Wullich, Robin and Fankhauser, Mattias and L{\"o}ffel, Kaspar and Schmidt, Thomas and El Kazzi, Mario},
  journal={chemrXiv},
  year={2025}
}

@article{ziesche2023multi,
  title={Multi-Dimensional Characterization of Battery Materials},
  author={Ziesche, Ralf F and Heenan, Thomas MM and Kumari, Pooja and Williams, Jarrod and Li, Weiqun and Curd, Matthew E and Burnett, Timothy L and Robinson, Ian and Brett, Dan JL and Ehrhardt, Matthias J and others},
  journal={Advanced Energy Materials},
  volume={13},
  number={23},
  pages={2300103},
  year={2023},
  publisher={Wiley Online Library}
}

@article{maffettone2023self,
  title={Self-driving multimodal studies at user facilities},
  author={Maffettone, Phillip M and Allan, Daniel B and Campbell, Stuart I and Carbone, Matthew R and Caswell, Thomas A and DeCost, Brian L and Gavrilov, Dmitri and Hanwell, Marcus D and Joress, Howie and Lynch, Joshua and others},
  journal={arXiv preprint arXiv:2301.09177},
  year={2023}
}

@article{yin2024integrated,
  title={Integrated edge-to-exascale workflow for real-time steering in neutron scattering experiments},
  author={Yin, Junqi and Reshniak, Viktor and Liu, Siyan and Zhang, Guannan and Wang, Xiaoping and Xiao, Zhongcan and Morgan, Zachary and Pawledzio, Sylwia and Proffen, Thomas and Hoffmann, Christina and others},
  journal={Structural Dynamics},
  volume={11},
  number={6},
  year={2024},
  publisher={AIP Publishing}
}

@article{corrao2025modular,
  title={A modular framework for collaborative human-AI, multi-modal and multi-beamline synchrotron experiments},
  author={Corrao, Adam A and Maffettone, Phillip M and Ravel, Bruce and Caswell, Thomas A and Joress, Howie and Wilkins, Stuart I and Olds, Daniel},
  journal={arXiv preprint arXiv:2509.22959},
  year={2025}
}

@article{mcdannald2022fly,
  title={On-the-fly autonomous control of neutron diffraction via physics-informed Bayesian active learning},
  author={McDannald, Austin and Frontzek, Matthias and Savici, Andrei T and Doucet, Mathieu and Rodriguez, Efrain E and Meuse, Kate and Opsahl-Ong, Jessica and Samarov, Daniel and Takeuchi, Ichiro and Ratcliff, William and others},
  journal={Applied Physics Reviews},
  volume={9},
  number={2},
  year={2022},
  publisher={AIP Publishing}
}

@article{yager2023autonomous,
  title={Autonomous x-ray scattering},
  author={Yager, Kevin G and Majewski, Pawel W and Noack, Marcus M and Fukuto, Masafumi},
  journal={Nanotechnology},
  volume={34},
  number={32},
  pages={322001},
  year={2023},
  publisher={IOP Publishing}
}

@article{szymanski2023adaptively,
  title={Adaptively driven X-ray diffraction guided by machine learning for autonomous phase identification},
  author={Szymanski, Nathan J and Bartel, Christopher J and Zeng, Yan and Diallo, Mouhamad and Kim, Haegyeom and Ceder, Gerbrand},
  journal={npj Computational Materials},
  volume={9},
  number={1},
  pages={31},
  year={2023},
  publisher={Nature Publishing Group UK London}
}

@article{pithan2023closing,
  title={Closing the loop: autonomous experiments enabled by machine-learning-based online data analysis in synchrotron beamline environments},
  author={Pithan, Linus and Starostin, Vladimir and Mare{\v{c}}ek, David and Petersdorf, Lukas and V{\"o}lter, Constantin and Munteanu, Valentin and Jankowski, Maciej and Konovalov, Oleg and Gerlach, Alexander and Hinderhofer, Alexander and others},
  journal={Synchrotron Radiation},
  volume={30},
  number={6},
  pages={1064--1075},
  year={2023},
  publisher={International Union of Crystallography}
}

@article{noack2019kriging,
  title={A kriging-based approach to autonomous experimentation with applications to x-ray scattering},
  author={Noack, Marcus M and Yager, Kevin G and Fukuto, Masafumi and Doerk, Gregory S and Li, Ruipeng and Sethian, James A},
  journal={Scientific reports},
  volume={9},
  number={1},
  pages={11809},
  year={2019},
  publisher={Nature Publishing Group UK London}
}

@article{adams2024human,
  title={Human-in-the-loop for Bayesian autonomous materials phase mapping},
  author={Adams, Felix and McDannald, Austin and Takeuchi, Ichiro and Kusne, A Gilad},
  journal={Matter},
  volume={7},
  number={2},
  pages={697--709},
  year={2024},
  publisher={Elsevier}
}

@article{li2020fourier,
  title={Fourier neural operator for parametric partial differential equations},
  author={Li, Zongyi and Kovachki, Nikola and Azizzadenesheli, Kamyar and Liu, Burigede and Bhattacharya, Kaushik and Stuart, Andrew and Anandkumar, Anima},
  journal={arXiv preprint arXiv:2010.08895},
  year={2020}
}

@article{du2024conditional,
  title={Conditional neural field latent diffusion model for generating spatiotemporal turbulence},
  author={Du, Pan and Parikh, Meet Hemant and Fan, Xiantao and Liu, Xin-Yang and Wang, Jian-Xun},
  journal={Nature Communications},
  volume={15},
  number={1},
  pages={10416},
  year={2024},
  publisher={Nature Publishing Group UK London}
}

@article{gao2024generative,
  title={Generative learning for forecasting the dynamics of high-dimensional complex systems},
  author={Gao, Han and Kaltenbach, Sebastian and Koumoutsakos, Petros},
  journal={Nature Communications},
  volume={15},
  number={1},
  pages={8904},
  year={2024},
  publisher={Nature Publishing Group UK London}
}

@article{de_castro_vargas_fernandes_absolute_2024,
	title = {Absolute permeability estimation from microtomography rock images through deep learning super-resolution and adversarial fine tuning},
	volume = {14},
	issn = {2045-2322},
	url = {https://www.ncbi.nlm.nih.gov/pmc/articles/PMC11271552/},
	doi = {10.1038/s41598-024-67367-1},
	pages = {16704},
    year = {2024},
	journal={Scientific Reports},
	shortjournal = {Sci Rep},
	author = {de Castro Vargas Fernandes, Júlio and Duarte Vidal, Alyne and Carvalho Medeiros, Lizianne and Menezes dos Anjos, Carlos Eduardo and Surmas, Rodrigo and Gonçalves Evsukoff, Alexandre},
	urldate = {2025-02-25},
	date = {2024-07-19},
}
